\documentclass[trackchanges,twocolumn,twocolappendix]{aastex701}

\usepackage{longtable}
\usepackage{paracol}
\usepackage{amsmath}

\newcommand{\ns}{}

\newcommand{\teff}{T_{\rm eff}}
\newcommand{\logg}{\log g}
\newcommand{\vmic}{\xi_{\rm t}}
\newcommand{\vmac}{V_{\rm mac}}
\newcommand{\vbrd}{v_{\rm brd}}

\begin{document}

\title{Observational constraints on the origin of the elements. X. Combining NLTE and machine learning for chemical diagnostics of 4 million stars in the 4MIDABLE-HR survey}

\author[orcid=0000-0002-5259-3974,sname='North America']{Nicholas Storm}
\affiliation{Max-Planck-Institut f\"{u}r Astronomie, K\"{o}nigstuhl 17, D-69117 Heidelberg, Germany}
\affiliation{Heidelberg University, Grabengasse 1, 69117 Heidelberg, Germany}
\email[show]{storm@mpia.de}

\author[orcid=0000-0002-9908-5571,sname='North America']{Maria Bergemann}
\affiliation{Max-Planck-Institut f\"{u}r Astronomie, K\"{o}nigstuhl 17, D-69117 Heidelberg, Germany}
\email[hide]{}  

\author[orcid=0000-0002-5819-3023,sname='North America']{Tomasz R\'o\.za\'nski}
\affiliation{Research School of Astronomy \& Astrophysics, 
             The Australian National University, Cotter Rd., Weston, ACT 2611, Australia}
\email[hide]{}  

\author[orcid=0000-0002-0294-799X,sname='North America']{Victor F.~Ksoll}
\affiliation{Universität Heidelberg, Zentrum für Astronomie, Institut für Theoretische Astrophysik, Albert-Ueberle-Str. 2, 69120 Heidelberg, Germany}
\email[hide]{}

\author[orcid=0000-0003-3978-1409,sname='North America']{Thomas Bensby}
\affiliation{Lund Observatory, Division of Astrophysics, Department of Physics, Lund University, Box 118, SE-221\,00 Lund, Sweden}
\email[hide]{}  

\author[orcid=0000-0002-1391-9097,sname='North America']{{Gregor Traven}}
\affiliation{Faculty of Mathematics and Physics, University of Ljubljana, Jadranska 19, 1000 Ljubljana, Slovenia}
\email[hide]{}  

\author[orcid=0000-0002-9035-3920,sname='North America']{{Georges Kordopatis}}
\affiliation{Université Côte d’Azur, Observatoire de la Côte d’Azur, CNRS, Laboratoire Lagrange, Nice, France}
\email[hide]{} 

\author[orcid=0000-0001-9204-0779,sname='North America']{Ross P. Church}
\affiliation{Lund Observatory, Division of Astrophysics, Department of Physics, Lund University, Box 118, SE-221\,00 Lund, Sweden}
\email[hide]{}  

\author[,sname='North America']{Mingjie Jian}
\affiliation{Institute of Astronomy, University of Cambridge, Madingley Road, Cambridge CB3 0HA, UK}
\email[hide]{}  

\author[orcid=0000-0002-3279-0233,sname='North America']{Weijia Sun}
\affiliation{Leibniz-Institut für Astrophysik Potsdam (AIP), An der Sternwarte 16, 14482 Potsdam, Germany}
\email[hide]{}  

\author[orcid=,sname='North America']{Guillaume Guiglion}
\affiliation{Zentrum f\"ur Astronomie der Universit\"at Heidelberg, Landessternwarte, K\"onigstuhl 12, 69117 Heidelberg, Germany}
\affiliation{Max-Planck-Institut f\"{u}r Astronomie, K\"{o}nigstuhl 17, D-69117 Heidelberg, Germany}
\email[hide]{}

\author[orcid=,sname='North America']{Gra\v{z}ina Tautvai\v{s}ien\.{e}}
\affiliation{Vilnius University, Faculty of Physics, Institute of Theoretical Physics and Astronomy, Sauletekio av. 3, 10257 Vilnius, Lithuania}
\email[hide]{}

\begin{abstract}

We present \ns{the} 4MOST-HR resolution Non-Local Thermal Equilibrium (NLTE) Payne artificial neural network (ANN), trained on $404\,793$ new FGK spectra with 16 elements computed in NLTE. This network will be part of the Stellar Abundances and atmospheric Parameters Pipeline (SAPP), which will analyse 4 million stars during the five year long 4MOST consortium 4: \ns{4MOST} MIlky way Disc And BuLgE High-Resolution (4MIDABLE-HR) survey. A fitting algorithm using this ANN is also presented that is able to fully-automatically and self-consistently derive both stellar parameters and elemental abundances. The ANN is validated by fitting 121 observed spectra of low-mass FGKM type stars, including main-sequence dwarf, subgiant and giant stars down to [Fe/H] \ns{$\approx -3.3$} degraded to 4MOST-HR resolution \ns{of $R\approx20\,000$}, and comparing the derived abundances with the output of the classical radiative transfer code TSFitPy. We are able to recover all 18 elemental abundances with a bias~$<0.13$ and spread~$<0.16$\,dex, although the typical values are $<0.09$ dex for most elements. These abundances are compared to the OMEGA+ Galactic Chemical Evolution model, showcasing for the first time, the expected performance and results obtained from high-resolution spectra of the quality expected to be obtained with 4MOST. The expected Galactic trends are recovered, and we highlight the potential of using many chemical elements to constrain the formation history of the Galaxy.

\end{abstract}

\keywords{\uat{Neural networks}{1933} --- \uat{Stellar abundances}{1577} --- \uat{Galaxy chemical evolution}{580}}

    \section{Introduction} 

Recent advances in spectroscopic facilities have enabled the collection of millions of individual stellar spectra in the Milky Way. One of them is the upcoming 4MOST facility with first light in late 2025. The 4MIDABLE-HR (4MOST consortium survey 4: MIlky way Disc And BuLgE High-Resolution) survey will provide high-resolution and high SNR spectra for over 4 million stars in the Galactic disc and the bulge during its five year program \citep{Bensby2019}. It will precisely measure chemical abundances not only to constrain their origin and production sites, but also to better understand the formation and history of our Galaxy \citep{Gilmore1995, BlandHawthorn2016, Barbuy2018, Cowan2021, Arcones2023}. However, analysis of this enormous amount of data cannot be efficiently done on a timely scale using classical spectral synthesis codes such as MOOG \citep{Sneden1973}, SIU \citep{Reetz1991}, PySME \citep{Wehrhahn2023}, KORG \citep{Wheeler2023}, or TSFitPy \citep{Gerber2023, Storm2023}. For this reason, a lot of spectroscopic surveys turned to different types of machine learning algorithms such as: the data-driven approach Cannon \citep{Ness2015}, convolutional neural networks (CNN) \citep[see e.g.][]{Guiglion2024b}, and the Payne \citep{Ting2019}, a forward generative neural network for synthetic spectra emulation. 

In this paper, we present an NLTE Payne artificial neural network (ANN) trained on $404~793$ newly computed spectra in non-local thermodynamic equilibrium (NLTE). Our main motivation is to provide a comprehensive and fast spectroscopic analysis code with state-of-the-art models computed - to the extent currently possible - using NLTE physics, which is required for accurate stellar parameter and chemical abundance analyses of stars \citep{Lind2024,Bergemann2025}. Our approach is based on a single network that emulates spectra with 16 elements simulated in NLTE across the full FGK parameter space. This emulator‑based approach provides fast spectral evaluations, introduces no explicit parameters priors, and integrates with standard fitting strategies that rely on features less sensitive to spectral‑model systematics (e.g. line‑list imperfections or blends). Moreover, emulation \ns{provides direct} access to the synthetic spectrum at the best‑fitting parameters, facilitating manual inspection of residuals and overall fit quality. While the NLTE Payne is not new \citep[e.g.][]{Kovalev2019, Buder2025}, \ns{we are not aware of any prior single‑network high resolution NLTE Payne ANN trained over such an extensive parameter space, with NLTE effects taken into account for 15+ elements. Although we note similar works by \citet{Buder2025}, which contained 14 elements in NLTE, but split the parameter space over many dozens of small neural networks; and works by \citet{Xiang2019, Zhang2025}, who utilised a slightly different architecture ``DD-Payne'' and applied it to lower resolution LAMOST spectra}. Our network is planned to be part of the Stellar Abundances and atmospheric Parameters Pipeline (SAPP) pipeline in 4MOST (for the 4MIDABLE-HR survey) and the PLATO missions \citep[][Lee et al. in prep.]{Gent2022}, but it can be used stand-alone as well. We also release this 4MOST-HR network for public use, together with 4MOST-LR and a high resolution version ($R=80~000$)\footnote{\url{https://nlte.mpia.de/gui-payne_fit.php}}, as well as the code used for Payne training\footnote{\url{https://github.com/stormnick/one_payne}}. \ns{We validate our new NLTE ANN abundances by comparing them with the results obtained using the classical radiative transfer code TSFitPy \citep{Gerber2023, Storm2023}. We also compare our self-consistent NLTE [X/Fe] trends with the predictions of our well-established Galactic Chemical Evolution (GCE) model OMEGA+ \citep{Cote2017, Cote2018b} to  highlight the parameter space of chemical enrichment studies, as planned with the 4MIDABLE-HR survey, but also to expose the current limitations and potential of such comparative analyses.}

The paper is organised as follows. We present our NLTE spectral synthetic models, trained NLTE Payne ANN, observed stellar sample and GCE model in Section~\ref{sec:data}. We validate the derived \ns{stellar parameters with literature and} abundances with results from classical fitting method with TSFitPy, and compare to the GCE models to constrain star formation history in Section~\ref{sec:results}. Finally we briefly summarise the conclusions in Section~\ref{sec:conclusions}.

\section{Data and Method}
\label{sec:data}

\subsection{NLTE Payne model}

We adopt \textit{The Payne} architecture \citep{Ting2019}, following its NLTE adaptation by \citet{Kovalev2019}. Recent scaling results \citep{2025OJAp....8E..69R} indicate that emulation accuracy improves predictably with training-set size, computing time spent for training, and neural network size; thus the achievable precision is set by resource and latency constraints. In this work, we require that a full per--spectrum fit (stellar parameters and 17 elemental abundances) \ns{is completed} within during around 10-20 seconds on a personal computer not equipped with GPU. \ns{This requirement is set such that we are able to fit spectra of several hundred (up to a several thousand) stars at high-resolution for each night of 4MOST observations.} Under this constraint, a compact Payne MLP offers a better accuracy-latency trade-off than heavier alternatives \citep[e.g. TransformerPayne;][]{Rozanski2025}, especially considering the fact that we can afford the computation of a spectral grid with hundreds of thousands of spectra. 

Our NLTE Payne is a four-layer multilayer perceptron (MLP). The input layer ingests 21 parameters: four \ns{atmospheric} stellar parameters \ns{(effective temperature $\teff$, surface gravity $\logg$, metallicity [Fe/H], and micro-turbulent velocity parameter $\vmic$}), and the abundances \ns{that are core for 4MIDABLE-HR survey science \citep[][Bergemann et al. in prep.]{Bensby2019}} of Li, C, O, Na, Mg, Al, Si, Ca, Ti, Cr, Mn, Co, Ni, Sr, Y, Ba, and Eu. Three fully connected hidden layers with 1024 units each and Sigmoid Linear Unit (SiLU) activations transform the inputs, and a final output layer, equipped with sigmoid activation \ns{function}, predicts continuum-normalised flux at 33\,375 wavelength pixels. This is the largest model that still satisfies our latency constraint. \ns{We have opted for SiLU in middle layers and for sigmoid in the output layer activation functions, because according to our tests, these functions results in lower final losses of the network. Further training details, the test results testing different activation functions} and hyper-parameter optimisation strategy are provided in Appendix~\ref{app:payne_training}.

\subsection{NLTE synthetic spectra}

\begin{figure}
\includegraphics[width=1\columnwidth]{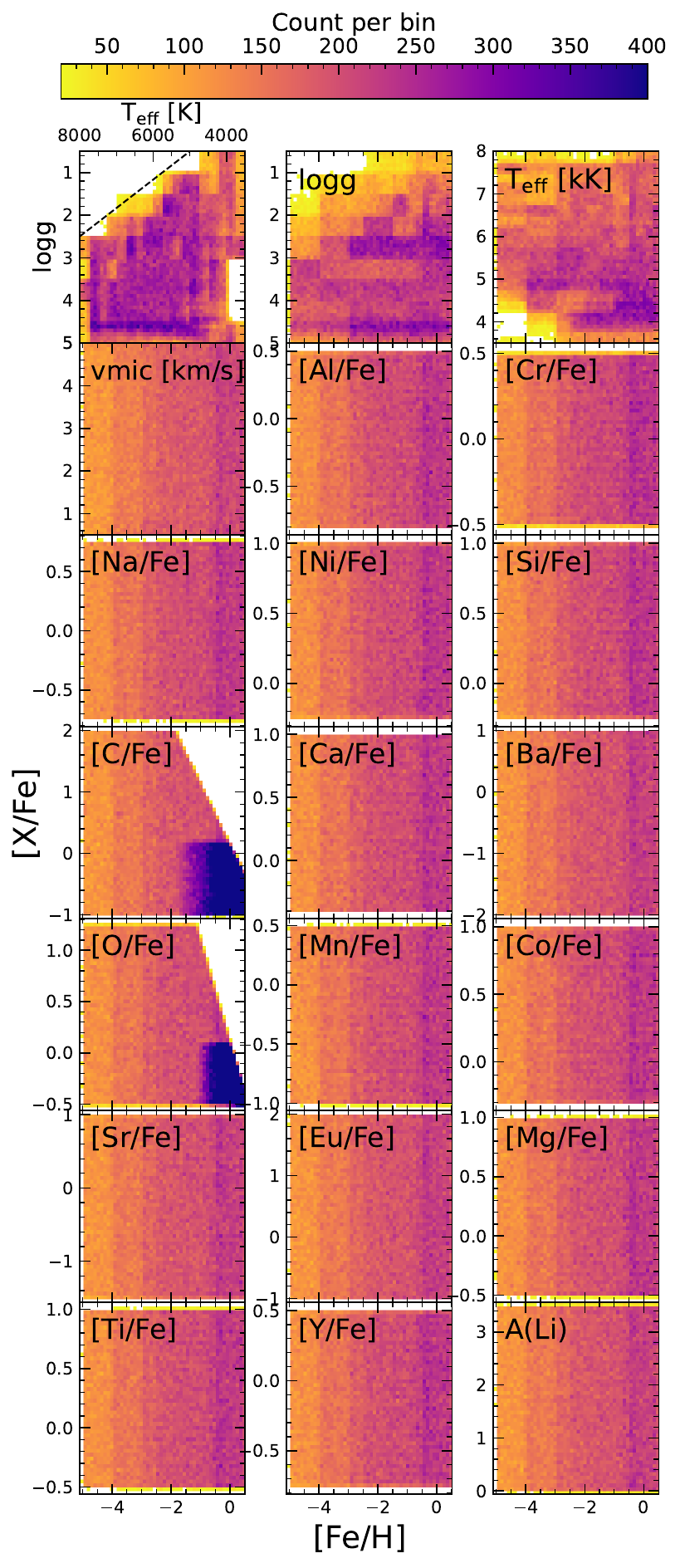}
\caption{Distribution in a 2D histogram of synthetic spectra used to train the Payne. Top left panel shows the Kiel diagram, while the rest are distributions as a function of [Fe/H]. All abundances are chosen uniformly random in metallicity space, except for A(O)$< 8.87$ and A(C)$< 8.7$. There are fewer low metallicity giant model atmospheres, resulting in slightly fewer spectra at low metallicities. There are also no public MARCS models above the black line in the $\teff$-$\logg$ space, resulting in a lack of computed spectra in that regime.
\label{fig:training_dist}}
\end{figure}

We trained the NLTE Payne described above on large, newly computed NLTE synthetic spectra grid with 404\,793 spectra, with generation requiring roughly \(125\,000\) CPU hours (see Fig.~\ref{fig:training_dist} for coverage), which also makes it one of the largest grids ever computed. The grid covers the 4MOST--HR blue (3926--4355~\AA), green (5160--5730~\AA), and red (6100--6790~\AA) windows \citep{deJong2019}. We created these spectra using the models and methods developed in our previous work \citep{Bergemann2019, Bergemann2021, Magg2022, Gerber2023}. We used the \ns{standard} 1D MARCS model atmospheres \citep{Gustafsson2008} and NLTE departure grids computed using \ns{our well-tested and validated} NLTE atomic models of H \citep{Mashonkina2008}, O \citep{Bergemann2021}, Na \citep{Ezzeddine2018}, Mg \citep{Bergemann2017}, Al \citep{Ezzeddine2018}, Si \citep{Bergemann2013, Magg2022}, Ca \citep{Mashonkina2017, Semenova2020}, Ti \citep{Bergemann2011}, Mn \citep{Bergemann2019}, Fe \citep{Bergemann2012a, Semenova2020}, Co \citep{Bergemann2010b, Yakovleva2020}, Ni \citep{Bergemann2021, Voronov2022}, Sr \citep{Bergemann2012b, Gerber2023}, Y \citep{Storm2023, Storm2024}, Ba \citep{Gallagher2020}, and Eu \citep{Storm2024}. \ns{The MARCS atmosphere grid relies on the chemical composition adopted from \citet{Grevesse1998, Grevesse2007}. However, in our NLTE synthetic spectra calculations we adopt our recent self-consistence solar composition from \citet{Magg2022}, which was obtained using Turbospectrum v.20 in NLTE. In App. \ref{app:solar_spectra} we demonstrate how the new NLTE Payne performs on the 4MOSTified solar spectra, and we then use the new solar abundances in the analysis of GCE results in Sec. \ref{sec:gce_model}.} 

Our grid coverage is presented in Fig. \ref{fig:training_dist} and covers the following ranges: $3500 \leq \teff  / \textrm{K} \leq 8000$, $0.5 \leq \logg \leq 5.0$, $-5 \leq \textrm{[Fe/H]} \leq 0.5$, $0.5 \leq \vmic  / \textrm{km s}^{-1} \leq 5$. Individual elements have distinct ranges that are slightly larger than the typically observed values. For red giants, there are no MARCS models at low metallicity, resulting in less spectra in that parameter space. Most importantly, the individual abundances were chosen uniformly random - there is no training bias (i.e. expected abundance pattern) that goes into the ANN. Only for C and O we chose values such that A(O)$< 8.87$ and A(C)$< 8.7$ to remove any spectra with very high abundance of C and O at high metallicity, causing unrealistic and too strong molecular bands in the training set. This also means that our network can effectively measure abundance in non-standard chemical composition stars. We adopted the Gaia-ESO line lists \citep{Heiter2021} with new atomic data for C, N, O, Si, Mg, as described in \citep{Magg2022}, and used VALD for its gaps \citep{Ryabchikova2015}. We also adopt solar abundances from \citet{Magg2022}. The NLTE Turbospectrum and TSFitPy wrapper \citep{Gerber2023, Storm2023} were employed to compute the spectra at infinite resolving power, which were further degraded to $R\approx20~000$ (``4MOSTified'') using the 4FS ETC package\footnote{\url{https://escience.aip.de/readthedocs/OpSys/etc/master/index.html}}. \ns{In order to simplify the Payne architecture, we have opted to not apply any macroscopic  broadening to the synthetic templates. Instead, we apply the convolution with a broadening profile after the neural network inference and during the $\chi^2$ minimisation parameter optimisation. Even at the resolving power of the 4MOST high-resolution spectrograph, it is not possible to disentangle different components, such as $\vmac$ and $V_{\rm sini}$, from the shape of spectral lines, which is why we only add a convolution with a the latter profile\footnote{See here: \url{https://github.com/stormnick/one_payne/blob/main/src/convolve.py}}. This implies that $\vbrd$ estimates represent the combined broadening due to projected (for unknown inclination) rotational velocity and macro-turbulence, but since the latter is an ad-hoc parameter introduced in 1D hydrostatic equilibrium models to compensate for the absence of realistic convection and turbulence \citep{Asplund2005, Nordlund2009, Lind2024}, we do not optimise this parameter further. This approach allows us to keep the Payne network loss low, while still allowing flexible fitting of rapidly rotating stars.} 
\subsection{Spectroscopic analysis}
Although the fitting methodology is not the main focus of this work, as it is similar to the classical synthetic codes, we briefly describe our method. We used \ns{the \texttt{minimize} function from Python package \texttt{LMFIT} \citep{Newville2025}} to fit the spectra, which takes on the order of 10-20 seconds to fit one spectrum (both stellar parameters and all 17 elemental abundances) on a single Mac M2 CPU core. 

The four \ns{atmospheric} parameters ($\teff$, $\logg$, [Fe/H], $\vmic$) and broadening ($\vbrd$) were determined by simultaneously fitting Fe~I, Fe~II, Mg~I and Ca~I lines \ns{({Table \ref{tab:lines}}). We did not impose priors on abundance ratios during the fit, neither did we adopt any scaling functions, such as e.g. the commonly-used $[\alpha/{\rm Fe}]$ relation with metallicity}. Fitting Fe-lines enforces ionisation balance, while Mg and Ca lines add additional constraints. We decided against usage of hydrogen lines, since they are badly modelled in giants, which gives poor results for those stars \citep[see e.g.][]{Wedemeyer2017}. The final best-fit values are then used to determine abundances \ns{by sequentially fitting abundances of remaining individual elements. Abundances of all 18 chemical elements (including Fe) were derived} by minimising the $\chi^2$ in the spectral regions contained by the line masks centred on the diagnostic lines. \ns{For Eu, we subtract 0.31 dex from all derived abundance, because in the standard linelist available to us, the $\log{gf}$ was offset by a constant factor $+0.31$ ($\log(2)$) compared to the accurate experimental f-value for this transition from \citet{Lawler2001}.} The list and atomic parameters of the fitted lines are provided in Tab. \ref{tab:lines}. \ns{Fe lines were chosen from the selection done based on \url{https://line-detector.oca.eu} \citep{Kordopatis2023b}.} However, we note that the line masks and the fitting algorithm are not fixed, and both can be easily adjusted without changing the NLTE ANN itself.

\subsection{Stellar sample}
\label{sec:stellar_sample}

\begin{figure}[ht!]
\includegraphics[width=1\columnwidth]{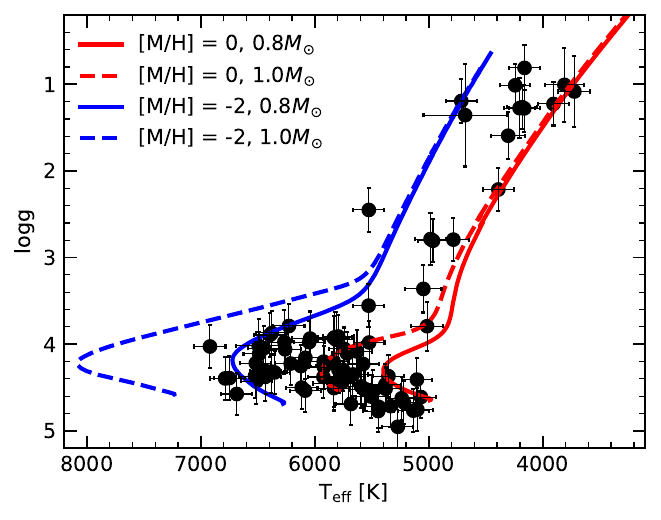}
\caption{Kiel diagram of the fitted stellar sample with [Fe/H] in colour with PARSEC evolutionary tracks \citep{Bressan2012} in colour.
\label{fig:hr_dia}}
\end{figure}

\begin{deluxetable*}{lcc}
\digitalasset
\tablewidth{0pt}
\tablecaption{Chosen parameters for the OMEGA+ GCE model. All other parameters were kept as default. \label{tab:omega_param}}
\tablehead{
\colhead{Parameter} & \colhead{Chosen/used values} & \colhead{Comment/yield source}
}
\startdata
Inflow rate & $\dot{M}_{\rm inflow}(t) = N_{\rm norm} e^{-t/\tau}$ & $N_{\rm norm} = 10 M_\odot$yr$^{-1}$, $\tau = 7\textrm{ Gyr}$ \\
$M_{DM}$& $10^{12} M_\odot$& constant at all times steps \\
SFE $\epsilon_\star$ & 0.03 & our default GCE model \\
Mass loading $\eta$& 1 & used in $\dot{M}_{\textrm{out}} = \eta \dot{M}_\star$ \\
NSM  & $\textrm{DTD}(\tau) \propto \tau^{-1}, 10\textrm{Myr} < \tau < 10^6\textrm{Gyr}$& \citet{arnould2007}\\
CCSN & 13-30 $M_\odot$ & \citet{Limongi2018}\\
MRSN & 0.01\% of 13-25 $M_\odot$ CCSN & \citet{Nishimura2015}\\
SNIa & 4 channels & \citet{Eitner2023}\\
AGB & 1-6 $M_\odot$ & \citet{Cristallo2015}\\
\enddata
\end{deluxetable*}

\begin{figure*}[ht!]
\includegraphics[width=0.93\textwidth]{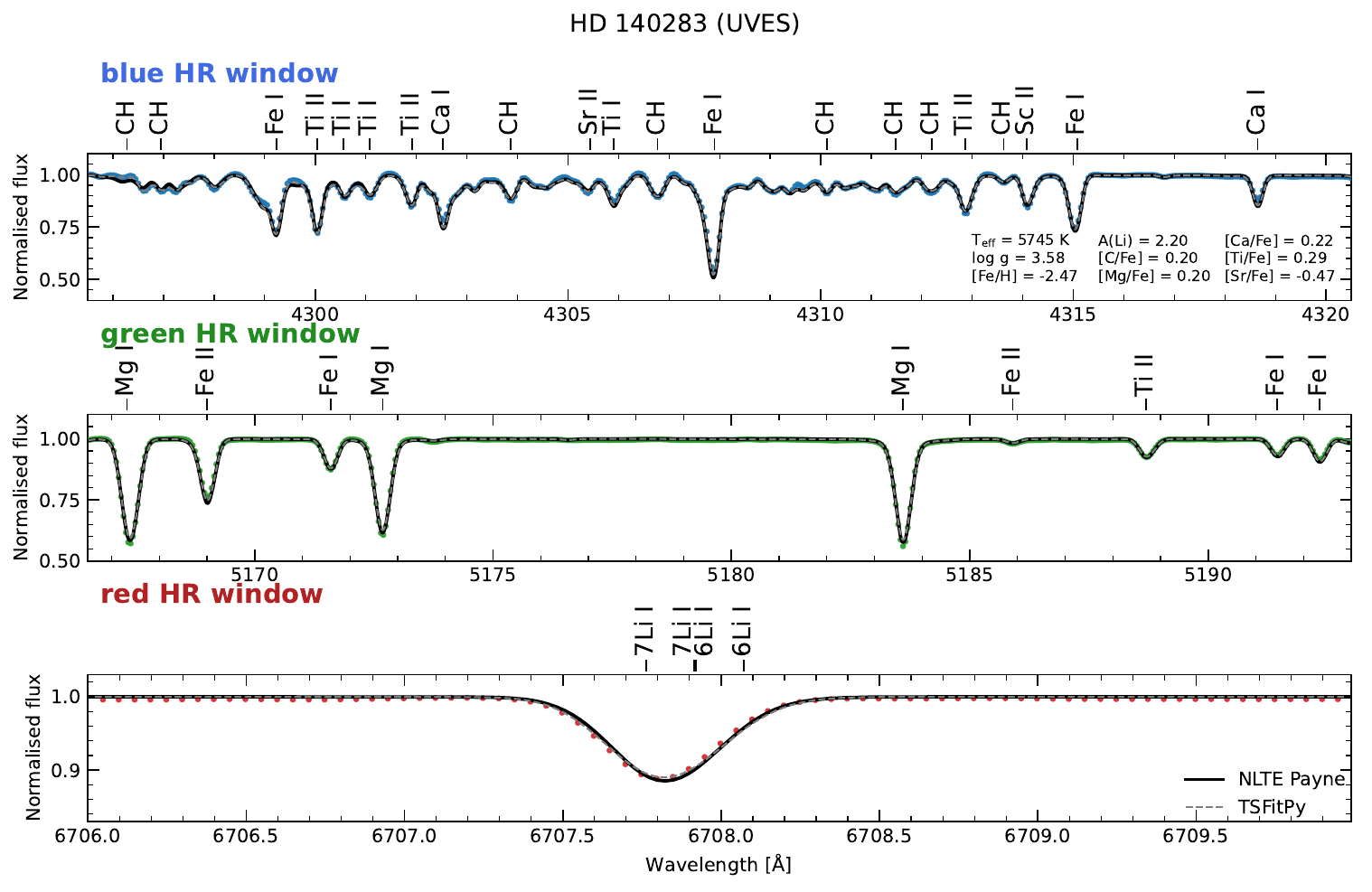}
\includegraphics[width=0.93\textwidth]{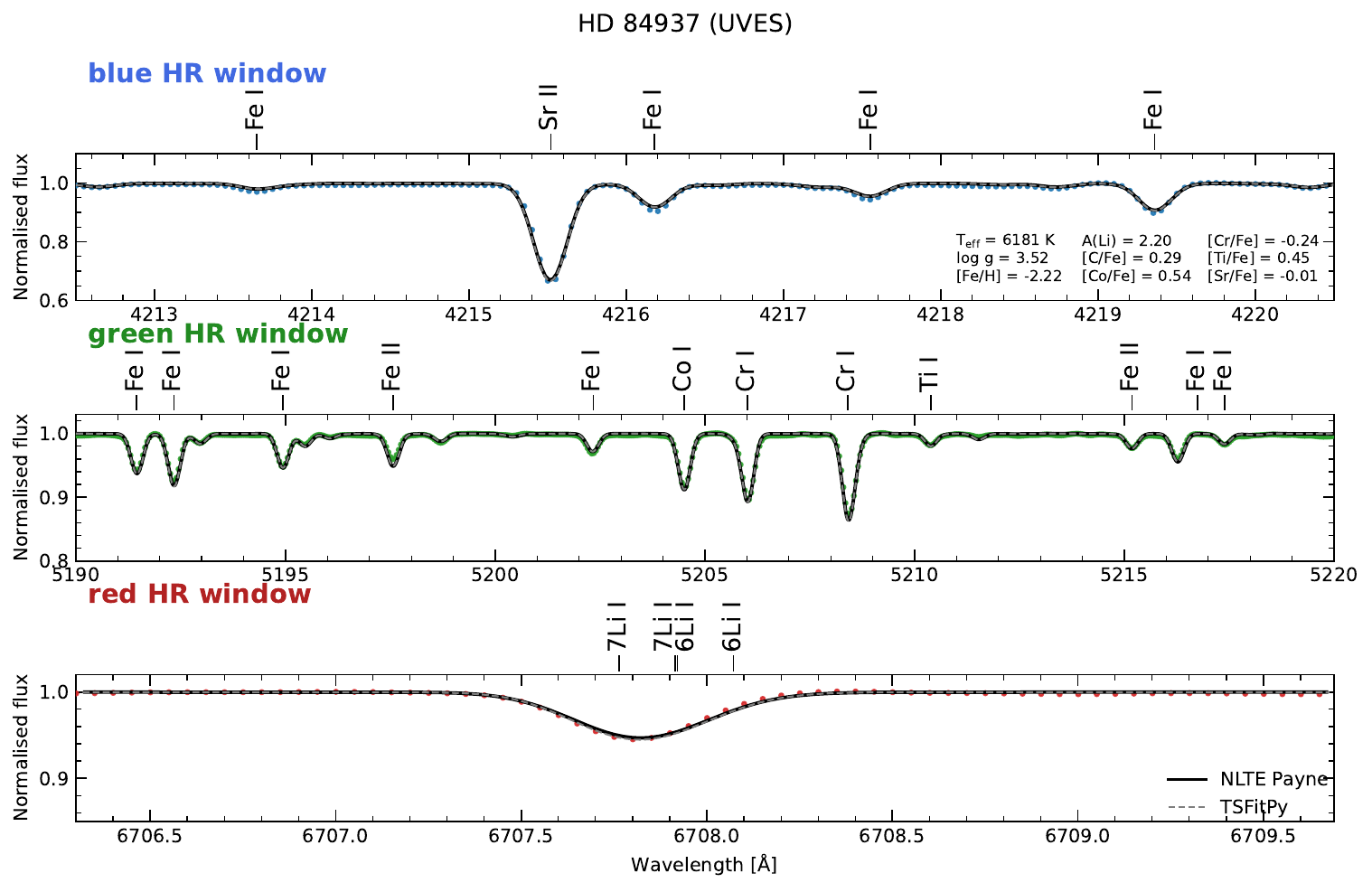}
\caption{\ns{NLTE Payne fit (black lines) to the HD 140283 and HD 84937 UVES spectra (dots), degraded to $R \approx 20000$ resolution, in all three 4MOST-HR windows. Emulated TSFitPy synthetic spectra (grey dashed line) using the best-fit abundances is overplotted and nearly perfectly matches the Payne fit.}
\label{fig:fit}}
\end{figure*}

Our calibration stars were selected from a sample of benchmark stars \citep{Heiter2015b}, and we also included spectra of nearby bright stars from our previous studies in \citet{Fuhrmann1993, Fuhrmann1998, Gehren2004, Gehren2006, Bergemann2008}, totalling 121 individual spectra \ns{for 84 stars}. The coverage of the stellar parameters is plotted in Fig. \ref{fig:hr_dia}, and covers the FGK\ns{M}-space in the ranges \ns{$3700 \lesssim \teff / \textrm{K} \lesssim 6900$, $0.8 \lesssim \logg \lesssim 5.0$ and $-3.35 \lesssim \textrm{[Fe/H]} \lesssim 0.34$. Our calibration sample consists of 48 main-sequence stars, 19 subgiants, and 17 red giants}. The observed spectra for the benchmark stars were taken from \citep{BlancoCuaresma2014} and they include high-resolution HARPS, NARVAL and UVES spectra, taken at an SNR of typically better than 200 per \AA. The wavelength coverage varies, and for the purpose of this work, we trimmed it to the 4MOST-HR windows. The rest of the sample was taken using FOCES, and the spectra have $R \approx 40\,000$, typically an SNR of at least 200 per \AA, and a wavelength range of 4500 to 6850 \AA~(with a few spectra extending from 3930 to 7450 \AA). \ns{Some of our spectra do not cover the blue 4MOST-HR window.}

We degraded the spectra to a resolution of $R \approx 20000$ using a simple FFT convolution. It is important to note that the actual 4MOST spectra are expected to have a resolution that varies with wavelength \ns{\citep[from $18\,000$ in the shorter-wavelength regime to $21\,000$ in the longer-wavelength regime for each spectral windows, see][]{deJong2019, Walcher2019}}\footnote{See also \url{www.4most.eu}}. Thus, we compared all derived Payne abundances with the ones fitted using TSFitPy. This is the only reliable way to do a self-consistent analysis and to showcase the capabilities of the NLTE ANN itself, without differences coming from using distinct spectra, model atmospheres, atomic data, spectra resolution etc., if compared to the literature values.

\subsection{GCE model}

To compute the chemical evolution of the elements discussed in this paper, we used the OMEGA+ GCE code \citep{Cote2017, Cote2018b}. This is a standard analytical GCE model and it tracks the galactic gas mass, increasing with the inflow from the circumgalactic medium (CGM) and the stellar mass-loss rate, and decreasing due to star formation and outflows. The code also tracks the elemental abundances in the interstellar medium (ISM) as a function of time, providing synthetic data in form of [X/Fe] that can be directly compared to observational data. The values of input parameters in OMEGA+ are provided in Tab. \ref{tab:omega_param}. The choices of these values are discussed and justified in \citet{Lian2023}, but we summarize them briefly here. We used an exponential inflow rate, which adds gas regardless of the available amount in CGM:

\begin{equation}
\label{eq:inflow}
    \dot{M}_{\rm inflow}(t) = N_{\rm norm} e^{-t/\tau},
\end{equation}

with $\dot{M}_{\rm inflow}$ is the inflow gas, $t$ age of the Galaxy in Gyr, $N_{\rm norm}$ is the normalisation constant (chosen as $10 M_\odot$yr$^{-1}$ here) and $\tau$ is infall timescale (chosen as 7 Gyr here). It is important point to note that OMEGA+ tracks the amount of gas in CGM ($M_{\rm CGM}$) with the initial amount at $t = 0$ defined as \citep[see also eq. 4 in][]{Cote2018b}:

\begin{equation}
    M_{\rm CGM}(t=0) = \frac{\Omega_{b,0}}{\Omega_0}M_{\rm vir},
\end{equation}

with adopted cosmological parameters $\Omega_{b,0} = 0.05$, $\Omega_0 = 0.32$, where $M_{\rm vir}$ includes both dark matter (DM) and baryonic mass (here initial $M_{\rm vir} \equiv M_{\rm DM}$, where $M_{\rm DM}$ is constant at all time steps). If at any time step the gas in CGM $M_{\rm CGM}$ runs out due to galactic inflows, then Eq.~\ref{eq:inflow} is not used during that time step for the inflow calculations; instead a small constant inflow rate is calculated:

\begin{equation}
\dot{M}_{\text{inflow}}(t) =
\begin{cases}
\dot{M}_{\text{calc}}, & \text{if } \dot{M}_{\text{calc}} \, \Delta t \le M_{\text{CGM}}, \\
M_{\text{CGM}} / \Delta t, & \text{if } \dot{M}_{\text{calc}} \, \Delta t > M_{\text{CGM}},
\end{cases}
\end{equation}

where $\Delta t$ is the size of the time step in years and $\dot{M}_{\text{calc}}$ is the calculated inflow rate from Eq. \ref{eq:inflow}. Therefore, if the initial CGM gas mass is too low or the infall rate is too large, then star formation will be quickly suppressed, as little new gas is added to the galaxy. However, as long as CGM has enough gas, the initial DM mass does not affect the inflow rate (which is the case for our galaxy simulation here). In OMEGA+, the star formation efficiency (SFE) $\epsilon_\star$ is a dimensionless parameter that goes into the equation:

\begin{equation}
    \dot{M}_\star = \frac{\epsilon_\star}{\tau_\star}M_{\text{gas}},
\end{equation}

where $\dot{M}_\star$ is the amount of gas going into the star formation, $M_{\text{gas}}$ is the amount of available gas and $\tau_\star$ is the star formation timescale. In our case, we keep the default configuration of $\tau_\star$, i.e. it is the dynamical timescale of the whole virialized system: 
\begin{equation}
    \tau_\star = 0.1 f_{\text{dyn}}H_0^{-1}(1+z)^{-3/2}, 
\end{equation}

where $H_0$ is the Hubble constant, $z$ is redshift and $f_{\text{dyn}} = 0.1$ by default \citep[see eq. 12 in][and references therein]{Cote2017}. $\tau_\star$ starts at $\approx6.7$ Myr and goes up to $\approx145.7$ Myr at the end of the simulation. In other words, varying the dimensionless SFE parameter $\epsilon_\star$ proportionally affects the timescale at which the star formation is happening.

Our GCE model relies on yields for core collapse supernovae (CCSN) from \citet[set 'R'][]{Limongi2018}, which uses yields from different rotations based on the velocity distribution weights derived by \citet{Prantzos2018}. For AGB stars we adopted the \citet{Cristallo2015} yields. We use the same approach as \citet{Eitner2023} for SNIa yields with four channels, each having its own delay time distribution (DTD): single degenerate Chandrasekhar mass SNIa with H-transfer from the companion via stable Roche lobe overflow, fainter SNeIax, sub-$M_{\textrm{ch}}$ with double-detonation of a C-O white dwarf (WD) and sub-$M_{\textrm{ch}}$ SNIa due to a merger of two WDs. For the r-process, we utilise two sites. The first are neutron star mergers (NSM) with a DTD slope $\tau^{-1}$, with the initial time delay of 10 Myr and maximum one until $10^6$ Gyr, according to the maximum merging time of NS-NS systems \citep{Beniamini2019}. The second are magnetorotational supernovae (MRSN), where we used the same approach as \citet{Kobayashi2020a}, by replacing 0.01\% of CCSN yields in the range 13-25 $M_\odot$ by yields of a 25 $M_\odot$ MRSN from \citet{Nishimura2015}.

\section{Results}
\label{sec:results}

\ns{In Fig. \ref{fig:fit} we show two examples of the NLTE ANN fit of a 4MOSTified spectra of the benchmark metal-poor stars HD~140283 and HD~84937.} Overall, the spectra are well reproduced. \ns{The Root Mean Squared Error (RMSE) of the flux is $\approx0.010$ in normalised flux units over all 3 windows for both spectra, which is fully sufficient for our purposes as the error is similar to the noise level of the data and other observational artefacts, such as data reduction and calibration, that introduce an even larger error.}

\subsection{Stellar parameter comparison}

\begin{figure}[ht!]
\includegraphics[width=1\columnwidth]{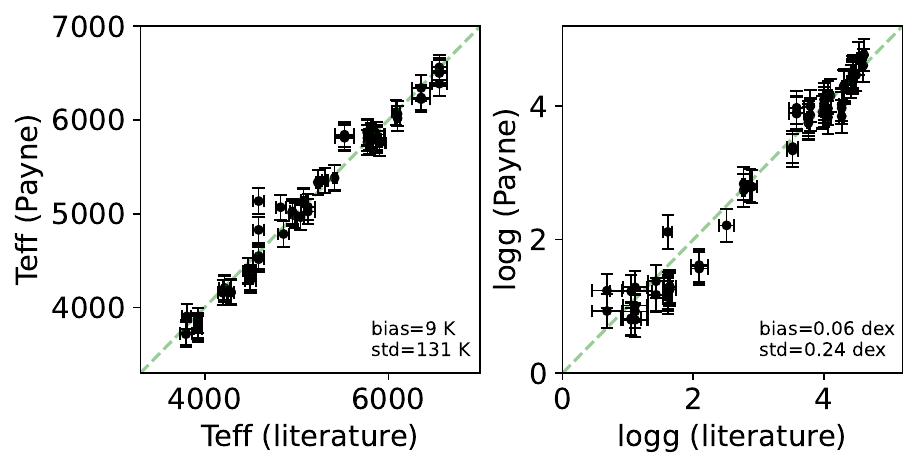}
\caption{\ns{Comparison between the literature stellar parameters of $\teff$ and $\logg$ for benchmark stars from \citet{Heiter2015b} to the Payne fitted ones.}
\label{fig:stellar_param_comparison}}
\end{figure}

\begin{figure*}[ht!]
\includegraphics[width=1\textwidth]{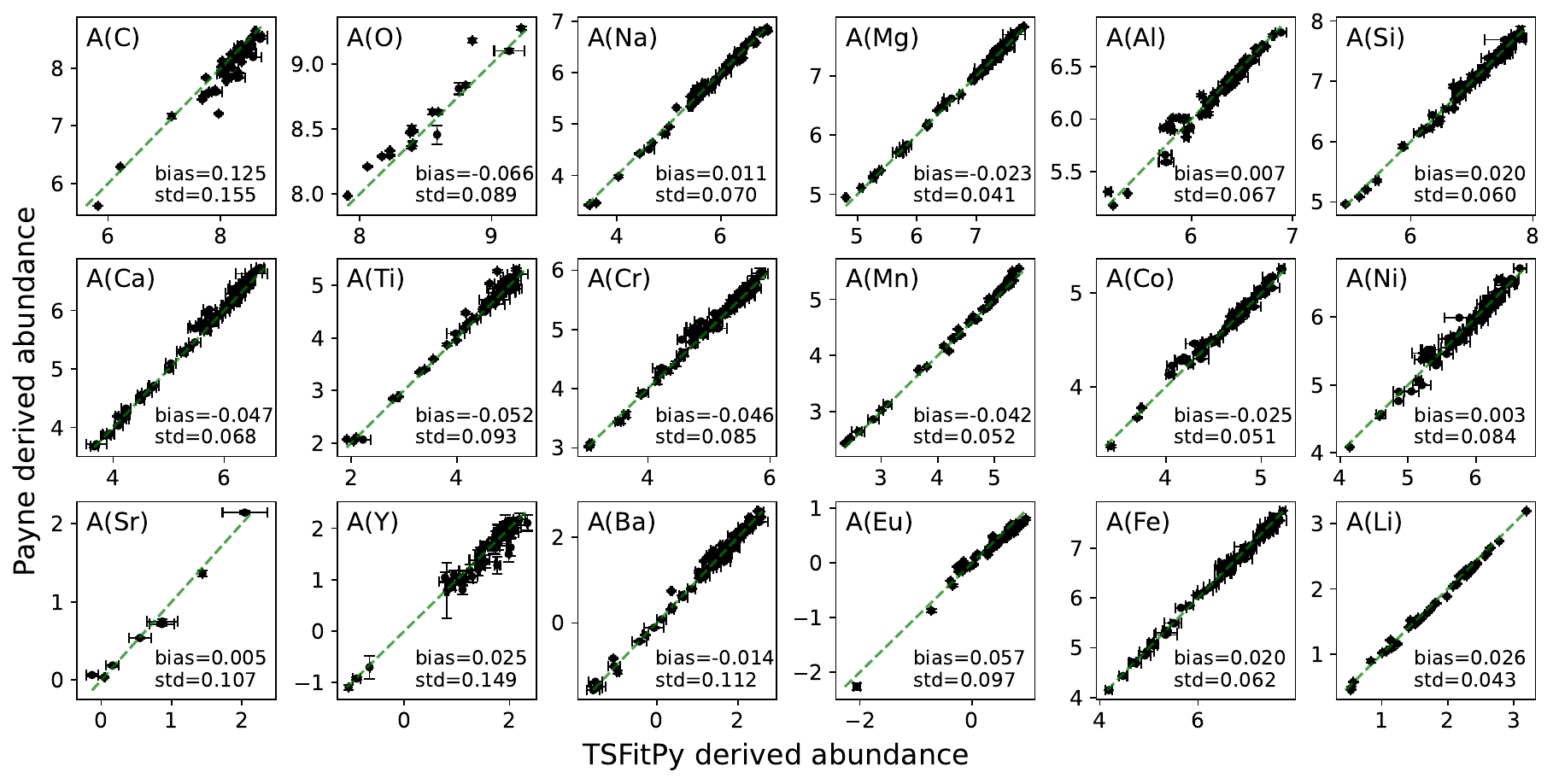}
\caption{Comparison between the derived abundance from Payne and TSFitPy plotted in absolute A(X) units. The green line is a one-to-one comparison. Each subpanel shows the average difference (bias) and standard deviation (std) when comparing the abundances from the two sources. 
\label{fig:ax_comparison}}
\end{figure*}

\begin{figure}[ht!]
\includegraphics[width=1\columnwidth]{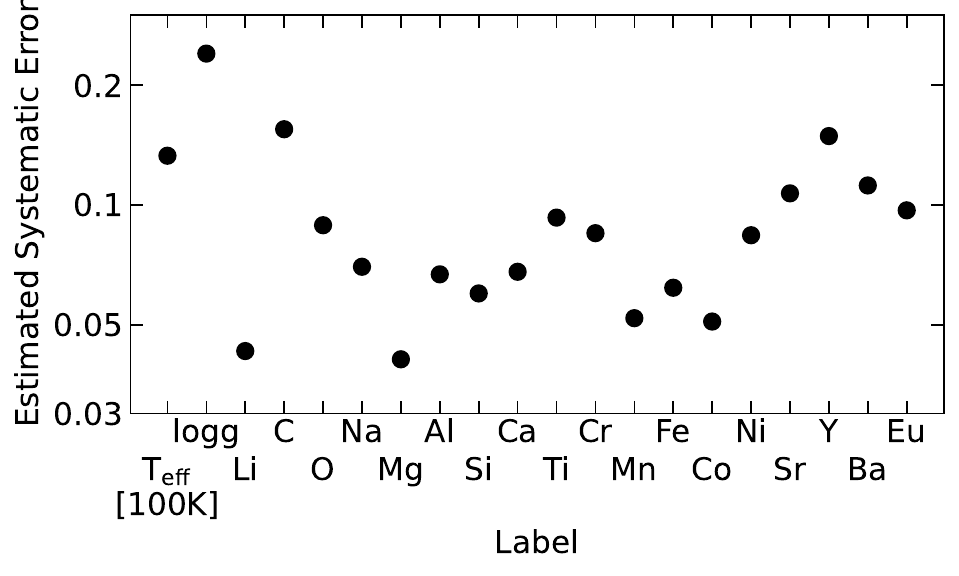}
\caption{Estimated systematic error for stellar parameters and abundances. The error was estimated by taking spread of the difference between the derived parameter from Payne and literature (for $\teff$ and $\logg$ for benchmark stars from \citet{Heiter2015b}) or TSFitPy. 
\label{fig:systematic_error}}
\end{figure}

\ns{We validated the stellar parameters $\teff$ and $\logg$ by comparing our best fit values to the values from the literature \citep{Heiter2015b}. This comparison has revealed a small systematic bias of $\approx -0.3$ dex in the spectroscopic $\logg$ values. This is not surprising: a similar bias was also noted in the analyses based on the previous NLTE Payne version \citep[see sect. 4.1 in][]{Gent2022}. This bias also affects the derived $\teff$ and [Fe/H] values, because of the degeneracy in the fitting of stellar parameters. The degeneracy can be broken only by using other non-spectroscopic information, such as photometric and parallax data \citep[e.g.][]{Schonrich2014, Buder2025}. The SAPP version in 4MOST will use this information \citep[see also][]{Serenelli2013, Schonrich2014}. Presently, in order to mitigate the bias in the current results, we apply a constant bias correction of $+0.3$ dex to $\logg$ and then we redetermine all other stellar parameters and abundances.}

\ns{Fig. \ref{fig:stellar_param_comparison} compares the literature stellar parameters with our NLTE ANN results for the benchmark stars. As mentioned above, the literature values were adopted from  \citet{Heiter2015b}, and these latter quantities rely on asteroseismic constraints for surface gravity and interferometric constraints on stellar angular diameter for $\teff$. The agreement between our values and the independent parameters is excellent; for $\teff$ and $\logg$, the average difference is $\sim$ 9K and 0.06~dex respectively. The standard deviation between the two distributions is 131~K and 0.24~dex. We note that using alternative sources of stellar parameters from \citet{Soubiran2024} results in similar differences: average differences of $\teff$ of 59~K and $\logg$ of 0.05~dex, and standard deviations of 105~K and 0.19~dex. This agreement is fully satisfactory, also given the fact that literature values are not free of systematic errors.}

\subsection{Abundance comparison}

The comparison between the absolute fitted abundance from Payne and TSFitPy is plotted in Fig. \ref{fig:ax_comparison}. Most spectra did not have blue window coverage, thus some elements were impossible to fit\ns{, especially} at low metallicity. For all elements, both bias and spread are respectively less than 0.13 and 0.16 dex, and $<0.09$ dex for most elements. Lithium, $\alpha$- and Fe-peak elements perform the best, given that they have the strongest and typically less-blended features. We mention some other elements below.

\ns{The star-to-star scatter of abundances of carbon (C) as derived using the NLTE Payne is much large compared to the values obtained the TSFitPy code. We trace this back to the lack of the CH G-band ($\approx4300$ \AA) in our observed 4MOSTified archival spectra. In the green, there is an atomic C I line at $5380$ \AA~and C$_2$ molecular features at 5164 and 5633 \AA, which are weak and they are typically not visible at [Fe/H] $\lesssim -1$. Because the 4MOST-HR spectra will have a much wider spectral range, covering the CH G-band, we do not expect such problems with the 4MIDABLE-HR data. The strongest atomic line of forbidden O at 6300 \AA~is severely blended with the telluric features, which in practice limits the use of this feature in the analysis of stellar O abundance \citep[][]{Pancino2017}. Some chemical elements only have few ($<5$) diagnostic lines, that may still be blended. Strontium has prominent diagnostic lines in the blue spectral range at 4077 (Sr I), 4161 (Sr I), 4215 (Sr II) and 4305 (Sr II) \AA. The Sr I line at 4077 \AA~is also blended with Fe I, which contributes to the uncertainty. Y II has only few lines in the green window and all of them are weak, resulting in a large spread of $\approx0.16$ dex. Also the key r-process element Eu II is represented by a weak line in the red window at 6645 \AA, typically used at higher [Fe/H], and there are two strong (yet blended) Eu II lines at 4129 and 4205 \AA~in the blue, which are only useful at lower metallicity. Therefore our estimates of stellar Eu abundances typically have a larger uncertainty of up to 0.15 dex.}

We estimate expected systematic errors by taking the standard deviation between the Payne and TSFitPy abundances. For $\teff$ and $\logg$, on the other hand, we use the differences with respect to  the \citet{Heiter2015b} results for the benchmark stars. Figure~\ref{fig:systematic_error} summarises both results. This systematic abundance error can be considered a conservative estimate, as the addition of the blue window would improve the determination for some elements. The stellar parameters estimates are also expected to improve, when derived as part of the full SAPP pipeline. Nevertheless, for all \ns{abundance} values we provide an error that is a flat sum of the systematic and statistical errors returned by the \ns{\texttt{LMFIT} \texttt{minimize} function. In the subsequent discussion (Sec. \ref{sec:gce_model}), we primarily focus on the abundances on the square bracket notation [X/Fe]. In these calculations, we use our self-consistent solar NLTE ANN abundances obtained using the 4MOSTified solar spectra \ref{tab:sun_corr}. Chemical abundances obtained from different spectra of the same star are averaged.}

\ns{In summary, we show that robust abundances are recovered by the NLTE Payne, and the values agree well with those obtained using the classical spectroscopic analysis methods.}

\subsection{Galactic Chemical Evolution Models}
\label{sec:gce_model}

\begin{figure*}[!htbp]
\includegraphics[width=1\textwidth]{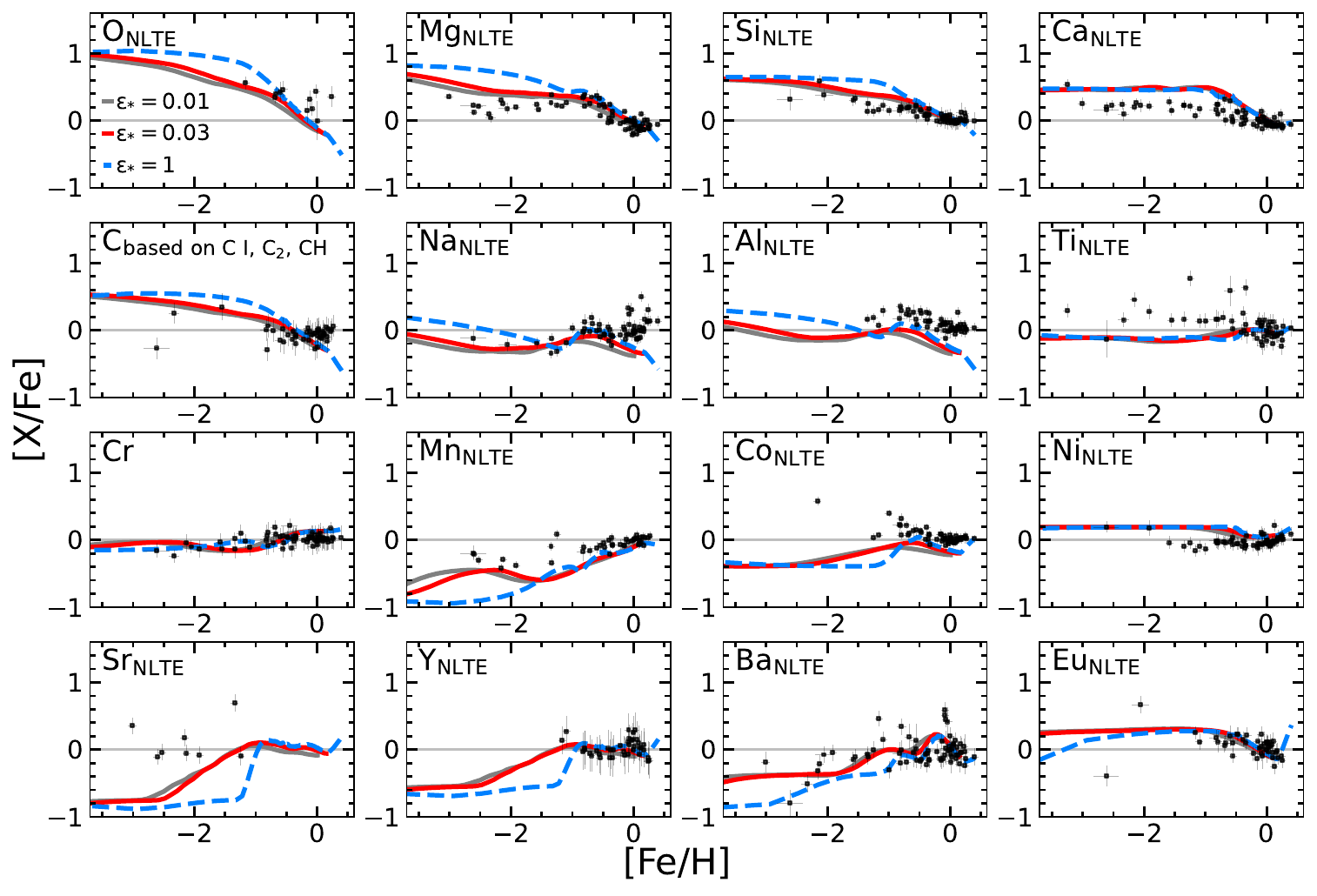}
\caption{Chemical abundances derived using the new NLTE Payne ANN (black points) from the observed archival data of Gaia-ESO benchmark and nearby bright stars. The archival spectra are described in Sect. \ref{sec:stellar_sample}. OMEGA+ GCE models are overplotted with three distinct star formation efficiencies $\epsilon_\star$: 0.01 (solid \ns{grey}), 0.03 (solid red) and 1 (dashed blue).}
\label{fig:gce}
\end{figure*}

\ns{Studies of detailed chemical abundance patterns enable probing quantitatively the Star Formation History (SFH) of a stellar population \citep[e.g.][]{Venn2004, Tolstoy2009, Bergemann2017, Matteucci2021}. Because stars of different mass have different lifetimes and their yields depend on mass and metallicity, the relative timing of CCSNe and SNe Ia leaves an imprint in the planes of $\alpha$ elements versus metallicity. But these differences are also readily visible in the chemical enrichment tracks of other chemical elements, enabling a more detailed set of constraints on the relative role of different events in the GCE. In this section, we focus solely on the Star Formation Efficiency (SFE) parameter as a proxy for the SFH - a key quantity in studies of galaxies \citep{Barbuy2018, Maiolino2019}, but we refer to \citet{Conroy2013, Somerville2015} for other types of constraints pertinent to Galaxy growth and formation history.}

\ns{Fig. \ref{fig:gce} shows our new NLTE Payne [X/Fe] abundances in comparison with the OMEGA$+$ GCE model computed using different values of SFE $\epsilon_\star$: 0.01, 0.03 \citep[our standard value, see][]{Lian2023} and 1. All other GCE parameters were kept at their default values as in our previous work \citep[see Tab. \ref{tab:omega_param} for the chosen parameters and also][]{Lian2023}. Following \citet{Hegedus2025}, we rescale the GCE trend for Mg by 0.3 dex upwards, in order to cure the misfit at low metallicity. For most chemical elements our [X/Fe] distributions are in excellent agreement with the literature using NLTE \citep{Zhao2016, Mishenina2017, Storm2025}. Especially the $\alpha$ elements, such as Mg, Si, Ti and Ca, follow the expected [Fe/H] - [$\alpha$/Fe] trends \citep{Bensby2014}. The higher value of the SFE shortens the chemical enrichment timescale, increasing the contribution of massive stars to the chemical evolution, which in turn implies that SNe Ia and AGBs start contributing to the enrichment at higher [Fe/H] (with a higher [Fe/H] achieved at a given time step due to a larger contribution of massive stars). As a consequence, the [$\alpha$/Fe] knee is shifted upwards. Adopting the SFE of 1 leads to higher - compared to observations - values of [Mg/Fe] and [Si/Fe] at low metallicity of [Fe/H] $\lesssim -2$. This difference potentially implies that lower SFE values would be preferred for the observed stellar population. Values of SFE lower than 0.03 do not yield any significant (here, under significant implying distinguishable given the uncertainty of the chemical abundance measurements) differences in the [Fe/H] - [X/Fe] plane, except for Mn at [Fe/H] $\lesssim -3$.} 

\ns{The proton (p)-rich elements \citep{Gehren2004, Gehren2006} Na and Al are produced in massive stars (in the Ne-Na and Mg-Al cycles) and in AGB stars \citep{Karakas2003, Tolstoy2009}. In our GCE model, an increased value of SFE leads to increased ratios of [Na/Fe] and [Al/Fe] (by $\sim 0.2-0.3$ dex) at [Fe/H] $\lesssim -1.5$. Also, the upturn of [Na/Fe] and [Al/Fe] as a function of [Fe/H] at [Fe/H] $\approx -1$ is due to the contribution from AGBs, as also seen in AGB models \citep{Forestini1997, Mowlavi1999}. The trends for [Na/Fe] and [Al/Fe] are similar to those presented by \citet{Bensby2014}, where they applied NLTE correction to Na abundances.}

\ns{The abundance ratios of iron-peak elements are higher compared to GCE model predictions, especially at low metallicity, which are consistent with our previous NLTE results for Mn, Co and Ni in 3D NLTE \citep{Storm2025}. The trend of [Mn/Fe] remains slightly sub-solar, whereas [Co/Fe] shows a mild increase towards lower metallicity, consistent with \citet{Bergemann2010b}. For Ni and Co, our observed sample is limited to [Fe/H] $\gtrsim -1.5$. For comparison, the LTE values of [Cr/Fe] and [Ni/Fe]  presented by \citet{Bensby2014} follow a similar flat trend with [Fe/H]. Similarly, the NLTE [Mn/Fe] and [Co/Fe] values of \citet{Battistini2015} resemble our results, however, we note that the latter paper relies on our older\footnote{These NLTE models made use of approximate classical rate coefficients describing the collisionally induced processes in H$+$Co and H$+$Mn transitions. In the later versions, these values were substituted by accurate quantum-mechanical values.} NLTE model atom of Mn and Co and therefore the direct comparison is not meaningful.}

\ns{Neutron-capture elements, such as Sr, Y and Ba, have significant r-process and weak s-process contributions from massive stars at the lowest metallicities. At higher metallicities on the other hand, the main s-process has a contribution only after the time delay of the formation of low and intermediate mass AGB stars. Thus, the metallicity at which the main s-process contributes gives an additional constraint on the SFH}. Our GCE model for the s-process element tracers [Sr, Y, Ba/Fe] $\lesssim 0$ at [Fe/H] $\lesssim -1.5$ exhibits a significant AGB rise at [Fe/H] $\gtrsim -2$. A SFE of 1 results in abundance ratios 0.2-0.6 dex lower compared to a higher SFE at a given metallicity for [Fe/H] $< -1$. Thus, the upturn in s-process offers insight into the onset of AGB enrichment and star formation timescale. \ns{Trends of ratios of [Y/Fe] and [Ba/Fe] reported by \citet{Bensby2014} are flat with [Fe/H], but some stars have supersolar [Ba/Fe] at solar [Fe/H]. We notice a similar pattern for our stars. For [Eu/Fe] we note an upturn towards the lower [Fe/H], similar to our GCE trend and the expected Galactic trend seen in \citet{Battistini2016}.} 

\ns{In summary, we show that the [X/Fe] obtained with the NLTE Payne are overall in agreement with results reported in the literature. Also, the comparison of [Fe/H] - [X/Fe] with the predictions of the OMEGA$+$ GCE model suggests that the abundances inferred from spectra are overall consistent with the expectations for the chemical enrichment of the Galactic disc. These NLTE trends provide a preview of the trends expected from 4MIDABLE-HR survey, which will be based on real 4MOST spectroscopic observations.}

\subsection{Lithium}

\begin{figure}[ht!]
\includegraphics[width=1\columnwidth]{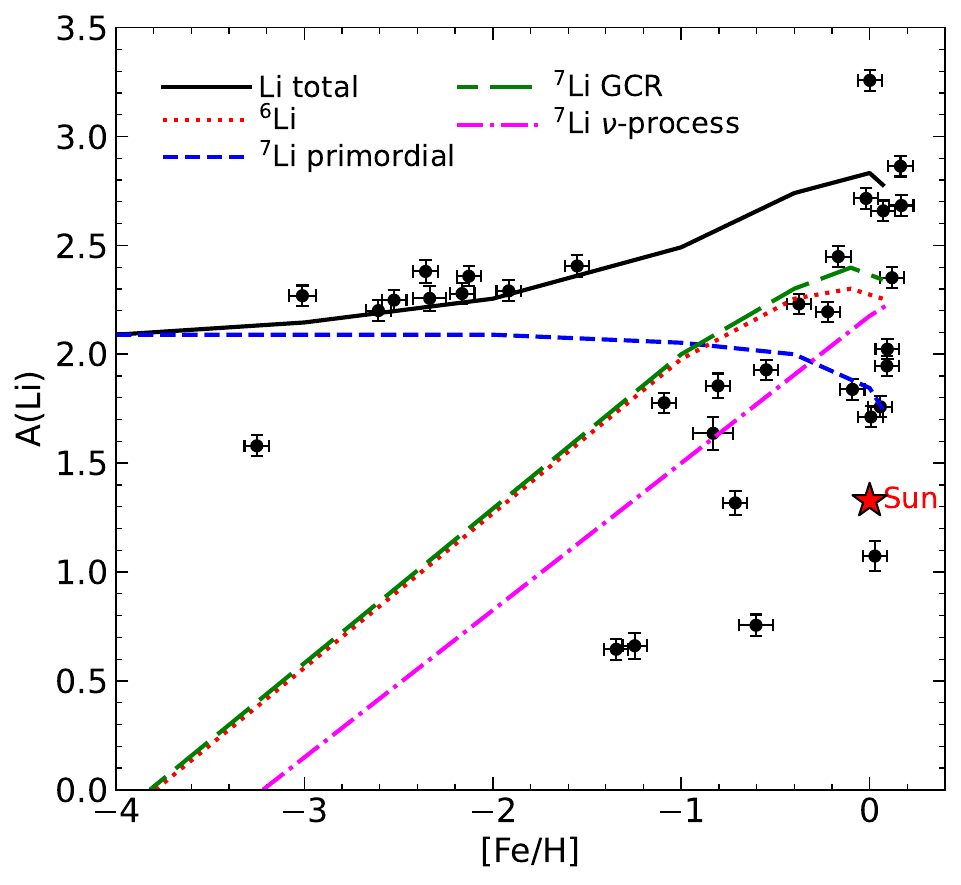}
\caption{Lithium abundance as a function of [Fe/H]. The GCE models from \citet{Fields1999a, Fields1999b} for different production channels are plotted, with the black line showing  the total value. The solar fitted value is marked as a red star.}
\label{fig:li}
\end{figure}

We discuss Li in a separate section, as we have not previously tested our OMEGA+ GCE model for Li. Instead we opt to use the \citet{Fields1999a, Fields1999b} models, which include several different production channels: primordial $^7$Li, production from Galactic cosmic-ray reactions (GCR), and production from the SNe $\nu$-process. 

Figure~\ref{fig:li} \ns{demonstrates} A(Li) as a function of [Fe/H], with our derived abundances \ns{shown} as black points and lines corresponding to different GCE \ns{distributions}. For [Fe/H] $\lesssim -1.6$ our stars lie on top of the curve with a scatter of $\approx 0.2$, which means that our derived Li abundance is consistent with the GCE model. \ns{Similar to the results of \citet{Ryan2000}, we find a large scatter of A(Li) at higher metallicities ([Fe/H] $\gtrsim -1.5$). They noted that the poor fit in this [Fe/H] regime may the consequence of missing production sources of Li, such as AGBs and classical novae \citep{Romano2021}. This is because each star would have its own production history, that cannot be captured by a unique GCE model. We note that the star with a very low Li abundances at [Fe/H] $= -3.25$ (HD 133442) is a cool giant ($\teff \approx 5520$ K), cooler than other metal-poor stars by $\approx 500$ K. Therefore the lower Li abundances is likely caused by destruction of it in the deeper convection envelope \citep[see similar discussions on this in e.g.][]{Spite1982}}. 

\section{Conclusions}
\label{sec:conclusions}

We present our new Non Local Thermodynamic Equilibrium (NLTE) astrophysical analysis code, which is based on the Payne artificial neural network. This code will be used \ns{within the SAPP pipeline in 4MOST Galactic Pipeline (4GP). Our code will} derive stellar parameters and abundances in 4MIDABLE-HR survey for over 4 million targets with the target SNR $> 100$ per \AA. To construct the new ANN, we computed a new grid of $404\,793$ NLTE synthetic spectral models, broadly covering the entire parameter space of FGK type stars and metallicities. \ns{Different architectures of the neural network, such as the activation functions, were carefully tested and adopted to give the lowest possible final training loss within the network requirements. We showed that SiLU activation functions halve the final loss, but also the ANN requires sigmoid activation function in the output layer for best performance.} The NLTE departure coefficients for the following 16 chemical elements are used: for H, O, Na, Mg, Al, Si, Ca, Ti, Mn, Fe, Co, Ni, Sr, Y, Ba and Eu \citep[see also][]{Gerber2023}. 

We release the new NLTE astrophysical analysis code at our group website\footnote{\url{https://nlte.mpia.de/gui-payne_fit.php}}. The online version of the code can be used to analyse 4MOST high-resolution stellar spectra in real-time. The line masks used are provided in Tab. \ref{tab:lines}. The code used for training is publicly available as well\footnote{\url{https://github.com/stormnick/one_payne}}. Our NLTE ANN code enables a direct and self-consistent analysis of spectra and determination of stellar parameters: $\teff$, $\logg$, [Fe/H], $\vmic$; broadening \ns{$\vbrd$}; and chemical abundances: A(Li), [C/Fe], [O/Fe], [Na/Fe], [Mg/Fe], [Al/Fe], [Si/Fe], [Ca/Fe], [Ti/Fe], [Cr/Fe], [Mn/Fe], [Co/Fe], [Ni/Fe], [Sr/Fe], [Y/Fe], [Ba/Fe], [Eu/Fe]. And it takes 10-20 seconds for a single spectrum on a single laptop CPU core to derive all stellar parameters and abundances.

We also validated the NLTE ANN performance on 67 unique archival spectra of Gaia-ESO benchmark stars and 54 additional spectra of nearby bright stars. The sample comprises of 84 stars, which cover the following parameter space \ns{$3700 \lesssim \teff / \textrm{K} \lesssim 6900$, $0.8 \lesssim \logg \lesssim 5.0$ and $-3.35 \lesssim \textrm{[Fe/H]} \lesssim 0.34$}. The observed spectra of these red giants, main-sequence stars, and subgiants were taken from HARPS, UVES, NARVAL and FOCES, and were degraded to the 4MOST-HR resolution $R \approx 20\,000$ to ensure consistency. For these stars, we also derive all chemical abundances using an independent code TSFitPy \citep{Gerber2023, Storm2023} and we show that the NLTE Payne machine learning based abundances are in excellent agreement with the values obtained using classical spectroscopic methods.

We furthermore compare our new NLTE abundances with
the Galactic chemical evolution models OMEGA+ \citep{Cote2017, Cote2018b}. In particular, we show how the star formation efficiency can be constrained using \ns{chemical elements such as Na, Al, O, Mg, Si, Sr, Ba, favouring lower values of the star formation efficiency, $\epsilon_\star \lesssim 0.03$ for the Galactic disc}. We show that the new NLTE Payne abundances demonstrate the same [X/Fe] behaviour as shown in previous NLTE studies, including O \citep{Amarsi2015b}, Na \citep{Zhao2016}, Mg \citep{Bergemann2017}, Al \citep{Zhao2016}, Si \citep{Zhao2016}, Ca \citep{Mashonkina2017}, Ti \citep{Bergemann2011}, Mn \citep{Eitner2020, Storm2025}, Ni \citep{Eitner2023, Storm2025}, Sr, Y, Ba, Eu \citep{Storm2025}. This work highlights the potential of 4MOST data in constraining the Galactic chemical enrichment and the origins of chemical elements comprehensively for the first time using the newly trained NLTE ANN.

\begin{acknowledgments}

We sincerely thank the referee for the positive feedback and comments that improved the readability of the paper.

\ns{Based on 4MOST observations collected at the European Southern Observatory under ESO programme 116.29EL.001. Funding for the construction of 4MOST, its science operations, and data processing has been provided by the 4MOST Consortium: https://4most.eu/cons.}

We thank the 4MOST IWG7 working group and Tadafumi Matsuno for valuable discussions. NS, MB and GG acknowledge funding from the European Research Council (ERC) under the European Union’s Horizon 2020 research and innovation programme (Grant agreement No. 949173). MB is supported through the Lise Meitner grant from the Max Planck Society. We acknowledge support by the Collaborative Research centre SFB 881 (projects A5, A10), Heidelberg University, of the Deutsche Forschungsgemeinschaft (DFG, German Research Foundation). TB acknowledges financial support by grant No. 2024-04990 from the Swedish Research Council. GG acknowledges support by Deutsche Forschungs-gemeinschaft (DFG, German Research Foundation) – project-IDs: eBer-22-59652 (GU 2240/1-1  ``Galactic Archaeology with Convolutional Neural-Networks: Realising the  potential of Gaia and 4MOST''). Computations were performed on the HPC systems Raven at the Max Planck Computing and Data Facility. VK acknowledges financial support from the German Excellence Strategy via the Heidelberg Cluster  ``STRUCTURES'' (EXC 2181 - 390900948), from the German Ministry for Economic Affairs and Climate Action in project ``MAINN'' (funding ID 50OO2206), and from the Carl Zeiss Stiftung. \ns{GK acknowledges support from the Bourse Qualité Recherche of the Observatoire de la Côte d'Azur (BQR OCA). GT acknowledges financial support of the Slovenian Research Agency (research core funding No. P1-0188) and the European Space Agency (Prodex Experiment Arrangement No. 4000143450).}

\end{acknowledgments}



\bibliography{sample701}{}

@ARTICLE{Bergemann2025,
       author = {{Bergemann}, Maria and {Hoppe}, Richard},
        title = "{3D Non-LTE radiation transfer: theory and applications to stars, exoplanets, and kilonovae}",
      journal = {arXiv e-prints},
         year = 2025,
        month = nov,
          eid = {arXiv:2511.04254},
        pages = {arXiv:2511.04254},
          doi = {10.48550/arXiv.2511.04254},
archivePrefix = {arXiv},
       eprint = {2511.04254},
 primaryClass = {astro-ph.SR},
       adsurl = {https://ui.adsabs.harvard.edu/abs/2025arXiv251104254B}
}

@ARTICLE{Lind2024,
       author = {{Lind}, Karin and {Amarsi}, Anish M.},
        title = "{Three-Dimensional Nonlocal Thermodynamic Equilibrium Abundance Analyses of Late-Type Stars}",
      journal = {\araa},
         year = 2024,
        month = sep,
       volume = {62},
       number = {1},
        pages = {475-527},
          doi = {10.1146/annurev-astro-052722-103557},
archivePrefix = {arXiv},
       eprint = {2401.00697},
 primaryClass = {astro-ph.SR},
       adsurl = {https://ui.adsabs.harvard.edu/abs/2024ARA&A..62..475L}
}

@ARTICLE{Voronov2022,
       author = {{Voronov}, Yaroslav V. and {Yakovleva}, Svetlana A. and {Belyaev}, Andrey K.},
        title = "{Inelastic Processes in Nickel-Hydrogen Collisions}",
      journal = {\apj},
         year = 2022,
        month = feb,
       volume = {926},
       number = {2},
          eid = {173},
        pages = {173},
          doi = {10.3847/1538-4357/ac46fd},
       adsurl = {https://ui.adsabs.harvard.edu/abs/2022ApJ...926..173V}
}

@ARTICLE{Lian2023,
       author = {{Lian}, Jianhui and {Storm}, Nicholas and {Guiglion}, Guillaume and {Serenelli}, Aldo and {Cote}, Benoit and {Karakas}, Amanda I. and {Boardman}, Nicholas and {Bergemann}, Maria},
        title = "{Observational constraints on the origin of the elements - VI. Origin and evolution of neutron-capture elements as probed by the Gaia-ESO survey}",
      journal = {\mnras},
         year = 2023,
        month = oct,
       volume = {525},
       number = {1},
        pages = {1329-1341},
          doi = {10.1093/mnras/stad2390},
archivePrefix = {arXiv},
       eprint = {2308.01111},
 primaryClass = {astro-ph.SR},
       adsurl = {https://ui.adsabs.harvard.edu/abs/2023MNRAS.525.1329L}
}

@ARTICLE{Battistini2015,
       author = {{Battistini}, Chiara and {Bensby}, Thomas},
        title = "{The origin and evolution of the odd-Z iron-peak elements Sc, V, Mn, and Co in the Milky Way stellar disk}",
      journal = {\aap},
         year = 2015,
        month = may,
       volume = {577},
          eid = {A9},
        pages = {A9},
          doi = {10.1051/0004-6361/201425327},
archivePrefix = {arXiv},
       eprint = {1502.01152},
 primaryClass = {astro-ph.GA},
       adsurl = {https://ui.adsabs.harvard.edu/abs/2015A&A...577A...9B}
}

@ARTICLE{Eitner2020,
       author = {{Eitner}, P. and {Bergemann}, M. and {Hansen}, C.~J. and {Cescutti}, G. and {Seitenzahl}, I.~R. and {Larsen}, S. and {Plez}, B.},
        title = "{Observational constraints on the origin of the elements. III. Evidence for the dominant role of sub-Chandrasekhar SN Ia in the chemical evolution of Mn and Fe in the Galaxy}",
      journal = {\aap},
         year = 2020,
        month = mar,
       volume = {635},
          eid = {A38},
        pages = {A38},
          doi = {10.1051/0004-6361/201936603},
archivePrefix = {arXiv},
       eprint = {2003.01721},
 primaryClass = {astro-ph.GA},
       adsurl = {https://ui.adsabs.harvard.edu/abs/2020A&A...635A..38E}
}

@ARTICLE{Eitner2023,
       author = {{Eitner}, P. and {Bergemann}, M. and {Ruiter}, A.~J. and {Avril}, O. and {Seitenzahl}, I.~R. and {Gent}, M.~R. and {C{\^o}t{\'e}}, B.},
        title = "{Observational constraints on the origin of the elements. V. NLTE abundance ratios of [Ni/Fe] in Galactic stars and enrichment by sub-Chandrasekhar mass supernovae}",
      journal = {\aap},
         year = 2023,
        month = sep,
       volume = {677},
          eid = {A151},
        pages = {A151},
          doi = {10.1051/0004-6361/202244286},
archivePrefix = {arXiv},
       eprint = {2206.10258},
 primaryClass = {astro-ph.GA},
       adsurl = {https://ui.adsabs.harvard.edu/abs/2023A&A...677A.151E}
}

@ARTICLE{Bergemann2010b,
       author = {{Bergemann}, Maria and {Pickering}, Juliet C. and {Gehren}, Thomas},
        title = "{NLTE analysis of CoI/CoII lines in spectra of cool stars with new laboratory hyperfine splitting constants}",
      journal = {\mnras},
         year = 2010,
        month = jan,
       volume = {401},
       number = {2},
        pages = {1334-1346},
          doi = {10.1111/j.1365-2966.2009.15736.x},
archivePrefix = {arXiv},
       eprint = {0909.2178},
 primaryClass = {astro-ph.SR},
       adsurl = {https://ui.adsabs.harvard.edu/abs/2010MNRAS.401.1334B}
}

@ARTICLE{Kovalev2019,
       author = {{Kovalev}, Mikhail and {Bergemann}, Maria and {Ting}, Yuan-Sen and {Rix}, Hans-Walter},
        title = "{Non-LTE chemical abundances in Galactic open and globular clusters}",
      journal = {\aap},
         year = 2019,
        month = aug,
       volume = {628},
          eid = {A54},
        pages = {A54},
          doi = {10.1051/0004-6361/201935861},
archivePrefix = {arXiv},
       eprint = {1907.02876},
 primaryClass = {astro-ph.SR},
       adsurl = {https://ui.adsabs.harvard.edu/abs/2019A&A...628A..54K}
}

@ARTICLE{Kobayashi2020a,
    author = {{Kobayashi}, Chiaki and {Karakas}, Amanda I. and {Lugaro}, Maria},
    title = "{The Origin of Elements from Carbon to Uranium}",
    journal = {\apj},
    year = 2020,
    month = sep,
    volume = {900},
    number = {2},
    eid = {179},
    pages = {179},
    doi = {10.3847/1538-4357/abae65},
    archivePrefix = {arXiv},
    eprint = {2008.04660},
    primaryClass = {astro-ph.GA},
    adsurl = {https://ui.adsabs.harvard.edu/abs/2020ApJ...900..179K}
}

@ARTICLE{Cowan2021,
    author = {{Cowan}, John J. and {Sneden}, Christopher and {Lawler}, James E. and {Aprahamian}, Ani and {Wiescher}, Michael and {Langanke}, Karlheinz and {Mart{\'\i}nez-Pinedo}, Gabriel and {Thielemann}, Friedrich-Karl},
    title = "{Origin of the heaviest elements: The rapid neutron-capture process}",
    journal = {Reviews of Modern Physics},
    year = 2021,
    month = jan,
    volume = {93},
    number = {1},
    eid = {015002},
    pages = {015002},
    doi = {10.1103/RevModPhys.93.015002},
    archivePrefix = {arXiv},
    eprint = {1901.01410},
    primaryClass = {astro-ph.HE},
    adsurl = {https://ui.adsabs.harvard.edu/abs/2021RvMP...93a5002C}
}

@ARTICLE{Gallagher2020,
       author = {{Gallagher}, A.~J. and {Bergemann}, M. and {Collet}, R. and {Plez}, B. and {Leenaarts}, J. and {Carlsson}, M. and {Yakovleva}, S.~A. and {Belyaev}, A.~K.},
        title = "{Observational constraints on the origin of the elements. II. 3D non-LTE formation of Ba II lines in the solar atmosphere}",
      journal = {\aap},
         year = 2020,
        month = feb,
       volume = {634},
          eid = {A55},
        pages = {A55},
          doi = {10.1051/0004-6361/201936104},
archivePrefix = {arXiv},
       eprint = {1910.03898},
 primaryClass = {astro-ph.SR},
       adsurl = {https://ui.adsabs.harvard.edu/abs/2020A&A...634A..55G}
}

@ARTICLE{Prantzos2018,
    author = {{Prantzos}, N. and {Abia}, C. and {Limongi}, M. and {Chieffi}, A. and {Cristallo}, S.},
    title = "{Chemical evolution with rotating massive star yields - I. The solar neighbourhood and the s-process elements}",
    journal = {\mnras},
    year = 2018,
    month = may,
    volume = {476},
    number = {3},
    pages = {3432-3459},
    doi = {10.1093/mnras/sty316},
    archivePrefix = {arXiv},
    eprint = {1802.02824},
    primaryClass = {astro-ph.GA},
    adsurl = {https://ui.adsabs.harvard.edu/abs/2018MNRAS.476.3432P}
}

@ARTICLE{Battistini2016,
    author = {{Battistini}, Chiara and {Bensby}, Thomas},
    title = "{The origin and evolution of r- and s-process elements in the Milky Way stellar disk}",
    journal = {\aap},
    year = 2016,
    month = feb,
    volume = {586},
    eid = {A49},
    pages = {A49},
    doi = {10.1051/0004-6361/201527385},
    archivePrefix = {arXiv},
    eprint = {1511.00966},
    primaryClass = {astro-ph.SR},
    adsurl = {https://ui.adsabs.harvard.edu/abs/2016A&A...586A..49B}
}

@ARTICLE{Zhao2016,
    author = {{Zhao}, G. and {Mashonkina}, L. and {Yan}, H.~L. and {Alexeeva}, S. and {Kobayashi}, C. and {Pakhomov}, Yu. and {Shi}, J.~R. and {Sitnova}, T. and {Tan}, K.~F. and {Zhang}, H.~W. and {Zhang}, J.~B. and {Zhou}, Z.~M. and {Bolte}, M. and {Chen}, Y.~Q. and {Li}, X. and {Liu}, F. and {Zhai}, M.},
    title = "{Systematic Non-LTE Study of the -.6 {\ensuremath{\leq}} [Fe/H] {\ensuremath{\leq}} 0.2 F and G Dwarfs in the Solar Neighborhood. II. Abundance Patterns from Li to Eu}",
    journal = {\apj},
    year = 2016,
    month = dec,
    volume = {833},
    number = {2},
    eid = {225},
    pages = {225},
    doi = {10.3847/1538-4357/833/2/225},
    archivePrefix = {arXiv},
    eprint = {1610.00193},
    primaryClass = {astro-ph.SR},
    adsurl = {https://ui.adsabs.harvard.edu/abs/2016ApJ...833..225Z}
}

@ARTICLE{Matteucci2021,
       author = {{Matteucci}, Francesca},
        title = "{Modelling the chemical evolution of the Milky Way}",
      journal = {\aapr},
         year = 2021,
        month = dec,
       volume = {29},
       number = {1},
          eid = {5},
        pages = {5},
          doi = {10.1007/s00159-021-00133-8},
archivePrefix = {arXiv},
       eprint = {2106.13145},
 primaryClass = {astro-ph.GA},
       adsurl = {https://ui.adsabs.harvard.edu/abs/2021A&ARv..29....5M}
}

@ARTICLE{Mashonkina2017,
    author = {{Mashonkina}, L. and {Jablonka}, P. and {Sitnova}, T. and {Pakhomov}, Yu. and {North}, P.},
    title = "{The formation of the Milky Way halo and its dwarf satellites; a NLTE-1D abundance analysis. II. Early chemical enrichment}",
    journal = {\aap},
    year = 2017,
    month = dec,
    volume = {608},
    eid = {A89},
    pages = {A89},
    doi = {10.1051/0004-6361/201731582},
    archivePrefix = {arXiv},
    eprint = {1709.04867},
    primaryClass = {astro-ph.SR},
    adsurl = {https://ui.adsabs.harvard.edu/abs/2017A&A...608A..89M}
}

@ARTICLE{Bensby2014,
    author = {{Bensby}, T. and {Feltzing}, S. and {Oey}, M.~S.},
    title = "{Exploring the Milky Way stellar disk. A detailed elemental abundance study of 714 F and G dwarf stars in the solar neighbourhood}",
    journal = {\aap},
    year = 2014,
    month = feb,
    volume = {562},
    eid = {A71},
    pages = {A71},
    doi = {10.1051/0004-6361/201322631},
    archivePrefix = {arXiv},
    eprint = {1309.2631},
    primaryClass = {astro-ph.GA},
    adsurl = {https://ui.adsabs.harvard.edu/abs/2014A&A...562A..71B}
}

@ARTICLE{Gent2022,
       author = {{Gent}, Matthew Raymond and {Bergemann}, Maria and {Serenelli}, Aldo and {Casagrande}, Luca and {Gerber}, Jeffrey M. and {Heiter}, Ulrike and {Kovalev}, Mikhail and {Morel}, Thierry and {Nardetto}, Nicolas and {Adibekyan}, Vardan and {Silva Aguirre}, V{\'\i}ctor and {Asplund}, Martin and {Belkacem}, Kevin and {del Burgo}, Carlos and {Bigot}, Lionel and {Chiavassa}, Andrea and {Rodr{\'\i}guez D{\'\i}az}, Luisa Fernanda and {Goupil}, Marie-Jo and {Gonz{\'a}lez Hern{\'a}ndezHanna}, Jonsay I. and {Mourard}, Denis and {Merle}, Thibault and {M{\'e}sz{\'a}ros}, Szabolcs and {Marshall}, Douglas J. and {Ouazzani}, Rhita-Maria and {Plez}, Bertrand and {Reese}, Daniel and {Trampedach}, Regner and {Tsantaki}, Maria},
        title = "{The SAPP pipeline for the determination of stellar abundances and atmospheric parameters of stars in the core program of the PLATO mission}",
      journal = {\aap},
         year = 2022,
        month = feb,
       volume = {658},
          eid = {A147},
        pages = {A147},
          doi = {10.1051/0004-6361/202140863},
archivePrefix = {arXiv},
       eprint = {2111.06666},
 primaryClass = {astro-ph.SR},
       adsurl = {https://ui.adsabs.harvard.edu/abs/2022A&A...658A.147G}
}

@ARTICLE{Bergemann2012a,
    author = {{Bergemann}, Maria and {Lind}, K. and {Collet}, R. and {Magic}, Z. and {Asplund}, M.},
    title = "{Non-LTE line formation of Fe in late-type stars - I. Standard stars with 1D and <3D> model atmospheres}",
    journal = {\mnras},
    year = 2012,
    month = nov,
    volume = {427},
    number = {1},
    pages = {27-49},
    doi = {10.1111/j.1365-2966.2012.21687.x},
    archivePrefix = {arXiv},
    eprint = {1207.2455},
    primaryClass = {astro-ph.SR},
    adsurl = {https://ui.adsabs.harvard.edu/abs/2012MNRAS.427...27B}
}

@ARTICLE{Bergemann2013,
       author = {{Bergemann}, Maria and {Kudritzki}, Rolf-Peter and {W{\"u}rl}, Matthias and {Plez}, Bertrand and {Davies}, Ben and {Gazak}, Zach},
        title = "{Red Supergiant Stars as Cosmic Abundance Probes. II. NLTE Effects in J-band Silicon Lines}",
      journal = {\apj},
         year = 2013,
        month = feb,
       volume = {764},
       number = {2},
          eid = {115},
        pages = {115},
          doi = {10.1088/0004-637X/764/2/115},
archivePrefix = {arXiv},
       eprint = {1212.2649},
 primaryClass = {astro-ph.SR},
       adsurl = {https://ui.adsabs.harvard.edu/abs/2013ApJ...764..115B}
}

@ARTICLE{Semenova2020,
    author = {{Semenova}, Ekaterina and {Bergemann}, Maria and {Deal}, Morgan and {Serenelli}, Aldo and {Hansen}, Camilla Juul and {Gallagher}, Andrew J. and {Bayo}, Amelia and {Bensby}, Thomas and {Bragaglia}, Angela and {Carraro}, Giovanni and {Morbidelli}, Lorenzo and {Pancino}, Elena and {Smiljanic}, Rodolfo},
    title = "{The Gaia-ESO survey: 3D NLTE abundances in the open cluster NGC 2420 suggest atomic diffusion and turbulent mixing are at the origin of chemical abundance variations}",
    journal = {\aap},
    year = 2020,
    month = nov,
    volume = {643},
    eid = {A164},
    pages = {A164},
    doi = {10.1051/0004-6361/202038833},
    archivePrefix = {arXiv},
    eprint = {2007.09153},
    primaryClass = {astro-ph.SR},
    adsurl = {https://ui.adsabs.harvard.edu/abs/2020A&A...643A.164S}
}

@ARTICLE{Bergemann2017,
       author = {{Bergemann}, Maria and {Collet}, Remo and {Amarsi}, Anish M. and {Kovalev}, Mikhail and {Ruchti}, Greg and {Magic}, Zazralt},
        title = "{Non-local Thermodynamic Equilibrium Stellar Spectroscopy with 1D and <3D> Models. I. Methods and Application to Magnesium Abundances in Standard Stars}",
      journal = {\apj},
         year = 2017,
        month = sep,
       volume = {847},
       number = {1},
          eid = {15},
        pages = {15},
          doi = {10.3847/1538-4357/aa88cb},
archivePrefix = {arXiv},
       eprint = {1612.07355},
 primaryClass = {astro-ph.SR},
       adsurl = {https://ui.adsabs.harvard.edu/abs/2017ApJ...847...15B}
}

@ARTICLE{Gerber2023,
       author = {{Gerber}, Jeffrey M. and {Magg}, Ekaterina and {Plez}, Bertrand and {Bergemann}, Maria and {Heiter}, Ulrike and {Olander}, Terese and {Hoppe}, Richard},
        title = "{Non-LTE radiative transfer with Turbospectrum}",
      journal = {\aap},
         year = 2023,
        month = jan,
       volume = {669},
          eid = {A43},
        pages = {A43},
          doi = {10.1051/0004-6361/202243673},
archivePrefix = {arXiv},
       eprint = {2206.00967},
 primaryClass = {astro-ph.SR},
       adsurl = {https://ui.adsabs.harvard.edu/abs/2023A&A...669A..43G}
}

@ARTICLE{arnould2007,
       author = {{Arnould}, M. and {Goriely}, S. and {Takahashi}, K.},
        title = "{The r-process of stellar nucleosynthesis: Astrophysics and nuclear physics achievements and mysteries}",
      journal = {\physrep},
         year = 2007,
        month = sep,
       volume = {450},
       number = {4-6},
        pages = {97-213},
          doi = {10.1016/j.physrep.2007.06.002},
archivePrefix = {arXiv},
       eprint = {0705.4512},
 primaryClass = {astro-ph},
       adsurl = {https://ui.adsabs.harvard.edu/abs/2007PhR...450...97A}
}

@ARTICLE{Bergemann2019,
       author = {{Bergemann}, Maria and {Gallagher}, Andrew J. and {Eitner}, Philipp and {Bautista}, Manuel and {Collet}, Remo and {Yakovleva}, Svetlana A. and {Mayriedl}, Anja and {Plez}, Bertrand and {Carlsson}, Mats and {Leenaarts}, Jorrit and {Belyaev}, Andrey K. and {Hansen}, Camilla},
        title = "{Observational constraints on the origin of the elements. I. 3D NLTE formation of Mn lines in late-type stars}",
      journal = {\aap},
         year = 2019,
        month = nov,
       volume = {631},
          eid = {A80},
        pages = {A80},
          doi = {10.1051/0004-6361/201935811},
archivePrefix = {arXiv},
       eprint = {1905.05200},
 primaryClass = {astro-ph.SR},
       adsurl = {https://ui.adsabs.harvard.edu/abs/2019A&A...631A..80B}
}

@ARTICLE{Gustafsson2008,
       author = {{Gustafsson}, B. and {Edvardsson}, B. and {Eriksson}, K. and {J{\o}rgensen}, U.~G. and {Nordlund}, {\r{A}}. and {Plez}, B.},
        title = "{A grid of MARCS model atmospheres for late-type stars. I. Methods and general properties}",
      journal = {\aap},
         year = 2008,
        month = aug,
       volume = {486},
       number = {3},
        pages = {951-970},
          doi = {10.1051/0004-6361:200809724},
archivePrefix = {arXiv},
       eprint = {0805.0554},
 primaryClass = {astro-ph},
       adsurl = {https://ui.adsabs.harvard.edu/abs/2008A&A...486..951G}
}

@ARTICLE{Nordlund2009,
       author = {{Nordlund}, {\r{A}}ke and {Stein}, Robert F. and {Asplund}, Martin},
        title = "{Solar Surface Convection}",
      journal = {Living Reviews in Solar Physics},
         year = 2009,
        month = dec,
       volume = {6},
       number = {1},
          eid = {2},
        pages = {2},
          doi = {10.12942/lrsp-2009-2},
       adsurl = {https://ui.adsabs.harvard.edu/abs/2009LRSP....6....2N}
}

@ARTICLE{Grevesse2007,
       author = {{Grevesse}, N. and {Asplund}, M. and {Sauval}, A.~J.},
        title = "{The Solar Chemical Composition}",
      journal = {\ssr},
         year = 2007,
        month = jun,
       volume = {130},
       number = {1-4},
        pages = {105-114},
          doi = {10.1007/s11214-007-9173-7},
       adsurl = {https://ui.adsabs.harvard.edu/abs/2007SSRv..130..105G}
}

@ARTICLE{Grevesse1998,
       author = {{Grevesse}, N. and {Sauval}, A.~J.},
        title = "{Standard Solar Composition}",
      journal = {\ssr},
         year = 1998,
        month = may,
       volume = {85},
        pages = {161-174},
          doi = {10.1023/A:1005161325181},
       adsurl = {https://ui.adsabs.harvard.edu/abs/1998SSRv...85..161G}
}

@ARTICLE{Maiolino2019,
       author = {{Maiolino}, R. and {Mannucci}, F.},
        title = "{De re metallica: the cosmic chemical evolution of galaxies}",
      journal = {\aapr},
         year = 2019,
        month = feb,
       volume = {27},
       number = {1},
          eid = {3},
        pages = {3},
          doi = {10.1007/s00159-018-0112-2},
archivePrefix = {arXiv},
       eprint = {1811.09642},
 primaryClass = {astro-ph.GA},
       adsurl = {https://ui.adsabs.harvard.edu/abs/2019A&ARv..27....3M}
}

@ARTICLE{Mishenina2017,
       author = {{Mishenina}, T. and {Pignatari}, M. and {C{\^o}t{\'e}}, B. and {Thielemann}, F.-K. and {Soubiran}, C. and {Basak}, N. and {Gorbaneva}, T. and {Korotin}, S.~A. and {Kovtyukh}, V.~V. and {Wehmeyer}, B. and {Bisterzo}, S. and {Travaglio}, C. and {Gibson}, B.~K. and {Jordan}, C. and {Paul}, A. and {Ritter}, C. and {Herwig}, F. and {NuGrid Collaboration}},
        title = "{Observing the metal-poor solar neighbourhood: a comparison of galactic chemical evolution predictions*{\textdagger}}",
      journal = {\mnras},
         year = 2017,
        month = aug,
       volume = {469},
       number = {4},
        pages = {4378-4399},
          doi = {10.1093/mnras/stx1145},
archivePrefix = {arXiv},
       eprint = {1705.03642},
 primaryClass = {astro-ph.SR},
       adsurl = {https://ui.adsabs.harvard.edu/abs/2017MNRAS.469.4378M}
}

@ARTICLE{Forestini1997,
       author = {{Forestini}, M. and {Charbonnel}, C.},
        title = "{Nucleosynthesis of light elements inside thermally pulsing AGB stars: I. The case of intermediate-mass stars}",
      journal = {\aaps},
         year = 1997,
        month = jun,
       volume = {123},
        pages = {241-272},
          doi = {10.1051/aas:1997348},
archivePrefix = {arXiv},
       eprint = {astro-ph/9608153},
 primaryClass = {astro-ph},
       adsurl = {https://ui.adsabs.harvard.edu/abs/1997A&AS..123..241F}
}

@ARTICLE{Mowlavi1999,
       author = {{Mowlavi}, Nami},
        title = "{Sodium production in asymptotic giant branch stars}",
      journal = {\aap},
         year = 1999,
        month = oct,
       volume = {350},
        pages = {73-88},
          doi = {10.48550/arXiv.astro-ph/9910542},
archivePrefix = {arXiv},
       eprint = {astro-ph/9910542},
 primaryClass = {astro-ph},
       adsurl = {https://ui.adsabs.harvard.edu/abs/1999A&A...350...73M}
}

@ARTICLE{Xiang2019,
       author = {{Xiang}, Maosheng and {Ting}, Yuan-Sen and {Rix}, Hans-Walter and {Sandford}, Nathan and {Buder}, Sven and {Lind}, Karin and {Liu}, Xiao-Wei and {Shi}, Jian-Rong and {Zhang}, Hua-Wei},
        title = "{Abundance Estimates for 16 Elements in 6 Million Stars from LAMOST DR5 Low-Resolution Spectra}",
      journal = {\apjs},
         year = 2019,
        month = dec,
       volume = {245},
       number = {2},
          eid = {34},
        pages = {34},
          doi = {10.3847/1538-4365/ab5364},
archivePrefix = {arXiv},
       eprint = {1908.09727},
 primaryClass = {astro-ph.SR},
       adsurl = {https://ui.adsabs.harvard.edu/abs/2019ApJS..245...34X}
}

@ARTICLE{Somerville2015,
       author = {{Somerville}, Rachel S. and {Dav{\'e}}, Romeel},
        title = "{Physical Models of Galaxy Formation in a Cosmological Framework}",
      journal = {\araa},
         year = 2015,
        month = aug,
       volume = {53},
        pages = {51-113},
          doi = {10.1146/annurev-astro-082812-140951},
archivePrefix = {arXiv},
       eprint = {1412.2712},
 primaryClass = {astro-ph.GA},
       adsurl = {https://ui.adsabs.harvard.edu/abs/2015ARA&A..53...51S}
}

@ARTICLE{Walcher2019,
       author = {{Walcher}, C.~J. and {Banerji}, M. and {Battistini}, C. and {Bell}, C.~P.~M. and {Bellido-Tirado}, O. and {Bensby}, T. and {Bestenlehner}, J.~M. and {Boller}, T. and {Brynnel}, J. and {Casey}, A. and {Chiappini}, C. and {Christlieb}, N. and {Church}, R. and {Cioni}, M.-R.~L. and {Croom}, S. and {Comparat}, J. and {Davies}, L.~J.~M. and {de Jong}, R.~S. and {Dwelly}, T. and {Enke}, H. and {Feltzing}, S. and {Feuillet}, D. and {Fouesneau}, M. and {Ford}, D. and {Frey}, S. and {Gonzalez-Solares}, E. and {Gueguen}, A. and {Howes}, L. and {Irwin}, M. and {Klar}, J. and {Kordopatis}, G. and {Korn}, A. and {Krumpe}, M. and {Kushniruk}, I. and {Lam}, M.~I. and {Lewis}, J. and {Lind}, K. and {Liske}, J. and {Loveday}, J. and {Mainieri}, V. and {Martell}, S. and {Matijevic}, G. and {McMahon}, R. and {Merloni}, A. and {Murphy}, D. and {Niederhofer}, F. and {Norberg}, P. and {Pramskiy}, A. and {Romaniello}, M. and {Robotham}, A.~S.~G. and {Rothmaier}, F. and {Ruchti}, G. and {Schnurr}, O. and {Schwope}, A. and {Smedley}, S. and {Sorce}, J. and {Starkenburg}, E. and {Stilz}, I. and {Storm}, J. and {Tempel}, E. and {Thi}, W.-F. and {Traven}, G. and {Valentini}, M. and {van den Ancker}, M. and {Walton}, N. and {Winkler}, R. and {Worley}, C.~C.},
        title = "{4MOST Scientific Operations}",
      journal = {The Messenger},
         year = 2019,
        month = mar,
       volume = {175},
        pages = {12-16},
          doi = {10.18727/0722-6691/5118},
archivePrefix = {arXiv},
       eprint = {1903.02465},
 primaryClass = {astro-ph.IM},
       adsurl = {https://ui.adsabs.harvard.edu/abs/2019Msngr.175...12W}
}

@software{Newville2025,
  author       = {Newville, Matthew and
                  Otten, Renee and
                  Nelson, Andrew and
                  Stensitzki, Till and
                  Ingargiola, Antonino and
                  Allan, Daniel and
                  Fox, Austin and
                  Carter, Faustin and
                  Rawlik, Michal},
  title        = {LMFIT: Non-Linear Least-Squares Minimization and
                   Curve-Fitting for Python
                  },
  month        = jul,
  year         = 2025,
  publisher    = {Zenodo},
  version      = {1.3.4},
  doi          = {10.5281/zenodo.16175987},
  url          = {https://doi.org/10.5281/zenodo.16175987},
  swhid        = {swh:1:dir:76742b0e41b1d2bff5a3716dd2376531f3a21db8
                   ;origin=https://doi.org/10.5281/zenodo.598352;visi
                   t=swh:1:snp:89f98ce93be53a573d85de0145f9174c827b7e
                   92;anchor=swh:1:rel:9528e5133d2d78c4036335f023b212
                   487cf1b632;path=lmfit-lmfit-py-0566445
                  },
}

@ARTICLE{Conroy2013,
       author = {{Conroy}, Charlie},
        title = "{Modeling the Panchromatic Spectral Energy Distributions of Galaxies}",
      journal = {\araa},
         year = 2013,
        month = aug,
       volume = {51},
       number = {1},
        pages = {393-455},
          doi = {10.1146/annurev-astro-082812-141017},
archivePrefix = {arXiv},
       eprint = {1301.7095},
 primaryClass = {astro-ph.CO},
       adsurl = {https://ui.adsabs.harvard.edu/abs/2013ARA&A..51..393C}
}

@ARTICLE{Zhang2025,
       author = {{Zhang}, Meng and {Xiang}, Maosheng and {Ting}, Yuan-Sen and {Amarsi}, Anish Mayur and {Zhang}, Hua-Wei and {Shi}, Jianrong and {Yuan}, Haibo and {Li}, Haining and {Wang}, Jiahui and {Wu}, Yaqian and {Wu}, Tianmin and {Mou}, Lanya and {Yan}, Hong-Liang and {Liu}, Jifeng},
        title = "{Homogeneous Stellar Atmospheric Parameters and 22 Elemental Abundances for FGK Stars Derived from LAMOST Low-resolution Spectra with DD-PAYNE}",
      journal = {\apjs},
         year = 2025,
        month = jul,
       volume = {279},
       number = {1},
          eid = {5},
        pages = {5},
          doi = {10.3847/1538-4365/add016},
archivePrefix = {arXiv},
       eprint = {2506.02763},
 primaryClass = {astro-ph.SR},
       adsurl = {https://ui.adsabs.harvard.edu/abs/2025ApJS..279....5Z}
}

@ARTICLE{Kordopatis2023b,
       author = {{Kordopatis}, Georges and {Hill}, Vanessa and {Lind}, Karin},
        title = "{Automatic line selection for abundance determinations in large stellar spectroscopic surveys}",
      journal = {\aap},
         year = 2023,
        month = jun,
       volume = {674},
          eid = {A104},
        pages = {A104},
          doi = {10.1051/0004-6361/202245684},
archivePrefix = {arXiv},
       eprint = {2302.11907},
 primaryClass = {astro-ph.IM},
       adsurl = {https://ui.adsabs.harvard.edu/abs/2023A&A...674A.104K}
}

@ARTICLE{Heiter2021,
       author = {{Heiter}, U. and {Lind}, K. and {Bergemann}, M. and {Asplund}, M. and {Mikolaitis}, {\v{S}}. and {Barklem}, P.~S. and {Masseron}, T. and {de Laverny}, P. and {Magrini}, L. and {Edvardsson}, B. and {J{\"o}nsson}, H. and {Pickering}, J.~C. and {Ryde}, N. and {Bayo Ar{\'a}n}, A. and {Bensby}, T. and {Casey}, A.~R. and {Feltzing}, S. and {Jofr{\'e}}, P. and {Korn}, A.~J. and {Pancino}, E. and {Damiani}, F. and {Lanzafame}, A. and {Lardo}, C. and {Monaco}, L. and {Morbidelli}, L. and {Smiljanic}, R. and {Worley}, C. and {Zaggia}, S. and {Randich}, S. and {Gilmore}, G.~F.},
        title = "{Atomic data for the Gaia-ESO Survey}",
      journal = {\aap},
         year = 2021,
        month = jan,
       volume = {645},
          eid = {A106},
        pages = {A106},
          doi = {10.1051/0004-6361/201936291},
archivePrefix = {arXiv},
       eprint = {2011.02049},
 primaryClass = {astro-ph.IM},
       adsurl = {https://ui.adsabs.harvard.edu/abs/2021A&A...645A.106H}
}

@ARTICLE{Yakovleva2020,
    author = {{Yakovleva}, Svetlana A. and {Belyaev}, Andrey K. and {Bergemann}, Maria},
    title = "{Cobalt-Hydrogen Atomic and Ionic Collisional Data}",
    journal = {Atoms},
    year = 2020,
    month = jul,
    volume = {8},
    number = {3},
    pages = {34},
    doi = {10.3390/atoms8030034},
    adsurl = {https://ui.adsabs.harvard.edu/abs/2020Atoms...8...34Y}
}

@ARTICLE{BlancoCuaresma2014,
       author = {{Blanco-Cuaresma}, S. and {Soubiran}, C. and {Jofr{\'e}}, P. and {Heiter}, U.},
        title = "{The Gaia FGK benchmark stars. High resolution spectral library}",
      journal = {\aap},
         year = 2014,
        month = jun,
       volume = {566},
          eid = {A98},
        pages = {A98},
          doi = {10.1051/0004-6361/201323153},
archivePrefix = {arXiv},
       eprint = {1403.3090},
 primaryClass = {astro-ph.SR},
       adsurl = {https://ui.adsabs.harvard.edu/abs/2014A&A...566A..98B}
}

@ARTICLE{Serenelli2013,
       author = {{Serenelli}, Aldo M. and {Bergemann}, Maria and {Ruchti}, Gregory and {Casagrande}, Luca},
        title = "{Bayesian analysis of ages, masses and distances to cool stars with non-LTE spectroscopic parameters}",
      journal = {\mnras},
         year = 2013,
        month = mar,
       volume = {429},
       number = {4},
        pages = {3645-3657},
          doi = {10.1093/mnras/sts648},
archivePrefix = {arXiv},
       eprint = {1212.4497},
 primaryClass = {astro-ph.SR},
       adsurl = {https://ui.adsabs.harvard.edu/abs/2013MNRAS.429.3645S}
}

@ARTICLE{Bergemann2012b,
    author = {{Bergemann}, M. and {Hansen}, C.~J. and {Bautista}, M. and {Ruchti}, G.},
    title = "{NLTE analysis of Sr lines in spectra of late-type stars with new R-matrix atomic data}",
    journal = {\aap},
    year = 2012,
    month = oct,
    volume = {546},
    eid = {A90},
    pages = {A90},
    doi = {10.1051/0004-6361/201219406},
    archivePrefix = {arXiv},
    eprint = {1207.2451},
    primaryClass = {astro-ph.SR},
    adsurl = {https://ui.adsabs.harvard.edu/abs/2012A&A...546A..90B}
}

@ARTICLE{Schonrich2014,
    author = {{Sch{\"o}nrich}, Ralph and {Bergemann}, Maria},
    title = "{Fundamental stellar parameters and metallicities from Bayesian spectroscopy: application to low- and high-resolution spectra}",
    journal = {\mnras},
    year = 2014,
    month = sep,
    volume = {443},
    number = {1},
    pages = {698-717},
    doi = {10.1093/mnras/stu1072},
    archivePrefix = {arXiv},
    eprint = {1311.5558},
    primaryClass = {astro-ph.SR},
    adsurl = {https://ui.adsabs.harvard.edu/abs/2014MNRAS.443..698S}
}

@ARTICLE{Magg2022,
       author = {{Magg}, Ekaterina and {Bergemann}, Maria and {Serenelli}, Aldo and {Bautista}, Manuel and {Plez}, Bertrand and {Heiter}, Ulrike and {Gerber}, Jeffrey M. and {Ludwig}, Hans-G{\"u}nter and {Basu}, Sarbani and {Ferguson}, Jason W. and {Gallego}, Helena Carvajal and {Gamrath}, S{\'e}bastien and {Palmeri}, Patrick and {Quinet}, Pascal},
        title = "{Observational constraints on the origin of the elements. IV. Standard composition of the Sun}",
      journal = {\aap},
         year = 2022,
        month = may,
       volume = {661},
          eid = {A140},
        pages = {A140},
          doi = {10.1051/0004-6361/202142971},
archivePrefix = {arXiv},
       eprint = {2203.02255},
 primaryClass = {astro-ph.SR},
       adsurl = {https://ui.adsabs.harvard.edu/abs/2022A&A...661A.140M}
}

@ARTICLE{Ryabchikova2015,
    author = {{Ryabchikova}, T. and {Piskunov}, N. and {Kurucz}, R.~L. and {Stempels}, H.~C. and {Heiter}, U. and {Pakhomov}, Yu and {Barklem}, P.~S.},
    title = "{A major upgrade of the VALD database}",
    journal = {\physscr},
    year = 2015,
    month = may,
    volume = {90},
    number = {5},
    eid = {054005},
    pages = {054005},
    doi = {10.1088/0031-8949/90/5/054005},
    adsurl = {https://ui.adsabs.harvard.edu/abs/2015PhyS...90e4005R}
}

@ARTICLE{Bensby2019,
       author = {{Bensby}, T. and {Bergemann}, M. and {Rybizki}, J. and {Lemasle}, B. and {Howes}, L. and {Kovalev}, M. and {Agertz}, O. and {Asplund}, M. and {Barklem}, P. and {Battistini}, C. and {Casagrande}, L. and {Chiappini}, C. and {Church}, R. and {Feltzing}, S. and {Ford}, D. and {Gerhard}, O. and {Kushniruk}, I. and {Kordopatis}, G. and {Lind}, K. and {Minchev}, I. and {McMillan}, P. and {Rix}, H. -W. and {Ryde}, N. and {Traven}, G.},
        title = "{4MOST Consortium Survey 4: Milky Way Disc and Bulge High-Resolution Survey (4MIDABLE-HR)}",
      journal = {The Messenger},
         year = 2019,
        month = mar,
       volume = {175},
        pages = {35-38},
          doi = {10.18727/0722-6691/5123},
archivePrefix = {arXiv},
       eprint = {1903.02470},
 primaryClass = {astro-ph.GA},
       adsurl = {https://ui.adsabs.harvard.edu/abs/2019Msngr.175...35B}
}

@ARTICLE{Asplund2005,
       author = {{Asplund}, Martin},
        title = "{New Light on Stellar Abundance Analyses: Departures from LTE and Homogeneity}",
      journal = {\araa},
         year = 2005,
        month = sep,
       volume = {43},
       number = {1},
        pages = {481-530},
          doi = {10.1146/annurev.astro.42.053102.134001},
       adsurl = {https://ui.adsabs.harvard.edu/abs/2005ARA&A..43..481A}
}

@ARTICLE{Bergemann2021,
       author = {{Bergemann}, Maria and {Hoppe}, Richard and {Semenova}, Ekaterina and {Carlsson}, Mats and {Yakovleva}, Svetlana A. and {Voronov}, Yaroslav V. and {Bautista}, Manuel and {Nemer}, Ahmad and {Belyaev}, Andrey K. and {Leenaarts}, Jorrit and {Mashonkina}, Lyudmila and {Reiners}, Ansgar and {Ellwarth}, Monika},
        title = "{Solar oxygen abundance}",
      journal = {\mnras},
         year = 2021,
        month = dec,
       volume = {508},
       number = {2},
        pages = {2236-2253},
          doi = {10.1093/mnras/stab2160},
archivePrefix = {arXiv},
       eprint = {2109.01143},
 primaryClass = {astro-ph.SR},
       adsurl = {https://ui.adsabs.harvard.edu/abs/2021MNRAS.508.2236B}
}

@ARTICLE{Lawler2001,
       author = {{Lawler}, J.~E. and {Wickliffe}, M.~E. and {den Hartog}, E.~A. and {Sneden}, C.},
        title = "{Improved Laboratory Transition Parameters forEu II and Application to the Solar Europium Elemental and Isotopic Composition}",
      journal = {\apj},
         year = 2001,
        month = dec,
       volume = {563},
       number = {2},
        pages = {1075-1088},
          doi = {10.1086/323407},
       adsurl = {https://ui.adsabs.harvard.edu/abs/2001ApJ...563.1075L}
}

@ARTICLE{Storm2023,
       author = {{Storm}, Nicholas and {Bergemann}, Maria},
        title = "{Observational constraints on the origin of the elements - VII. NLTE analysis of Y II lines in spectra of cool stars and implications for Y as a Galactic chemical clock}",
      journal = {\mnras},
         year = 2023,
        month = nov,
       volume = {525},
       number = {3},
        pages = {3718-3729},
          doi = {10.1093/mnras/stad2488},
archivePrefix = {arXiv},
       eprint = {2308.12092},
 primaryClass = {astro-ph.SR},
       adsurl = {https://ui.adsabs.harvard.edu/abs/2023MNRAS.525.3718S}
}

@ARTICLE{Mashonkina2008,
       author = {{Mashonkina}, L. and {Zhao}, G. and {Gehren}, T. and {Aoki}, W. and {Bergemann}, M. and {Noguchi}, K. and {Shi}, J.~R. and {Takada-Hidai}, M. and {Zhang}, H.~W.},
        title = "{Non-LTE line formation for heavy elements in four very metal-poor stars}",
      journal = {\aap},
         year = 2008,
        month = feb,
       volume = {478},
       number = {2},
        pages = {529-541},
          doi = {10.1051/0004-6361:20078060},
archivePrefix = {arXiv},
       eprint = {0711.4454},
 primaryClass = {astro-ph},
       adsurl = {https://ui.adsabs.harvard.edu/abs/2008A&A...478..529M}
}

@ARTICLE{Fuhrmann1993,
       author = {{Fuhrmann}, K. and {Axer}, M. and {Gehren}, T.},
        title = "{Balmer lines in cool dwarf stars. I. Basic influence of atmospheric models.}",
      journal = {\aap},
         year = 1993,
        month = apr,
       volume = {271},
        pages = {451-462},
       adsurl = {https://ui.adsabs.harvard.edu/abs/1993A&A...271..451F}
}

@ARTICLE{Storm2024,
       author = {{Storm}, N. and {Barklem}, P.~S. and {Yakovleva}, S.~A. and {Belyaev}, A.~K. and {Palmeri}, P. and {Quinet}, P. and {Lodders}, K. and {Bergemann}, M. and {Hoppe}, R.},
        title = "{3D NLTE modelling of Y and Eu. Centre-to-limb variation and solar abundances}",
      journal = {\aap},
         year = 2024,
        month = mar,
       volume = {683},
          eid = {A200},
        pages = {A200},
          doi = {10.1051/0004-6361/202348971},
archivePrefix = {arXiv},
       eprint = {2401.13450},
 primaryClass = {astro-ph.SR},
       adsurl = {https://ui.adsabs.harvard.edu/abs/2024A&A...683A.200S}
}

@ARTICLE{Ezzeddine2018,
       author = {{Ezzeddine}, R. and {Merle}, T. and {Plez}, B. and {Gebran}, M. and {Th{\'e}venin}, F. and {Van der Swaelmen}, M.},
        title = "{An empirical recipe for inelastic hydrogen-atom collisions in non-LTE calculations}",
      journal = {\aap},
         year = 2018,
        month = oct,
       volume = {618},
          eid = {A141},
        pages = {A141},
          doi = {10.1051/0004-6361/201630352},
       adsurl = {https://ui.adsabs.harvard.edu/abs/2018A&A...618A.141E}
}

@ARTICLE{Bergemann2011,
       author = {{Bergemann}, Maria},
        title = "{Ionization balance of Ti in the photospheres of the Sun and four late-type stars}",
      journal = {\mnras},
         year = 2011,
        month = may,
       volume = {413},
       number = {3},
        pages = {2184-2198},
          doi = {10.1111/j.1365-2966.2011.18295.x},
archivePrefix = {arXiv},
       eprint = {1101.0828},
 primaryClass = {astro-ph.SR},
       adsurl = {https://ui.adsabs.harvard.edu/abs/2011MNRAS.413.2184B}
}

@ARTICLE{Cote2017,
       author = {{C{\^o}t{\'e}}, Benoit and {O'Shea}, Brian W. and {Ritter}, Christian and {Herwig}, Falk and {Venn}, Kim A.},
        title = "{The Impact of Modeling Assumptions in Galactic Chemical Evolution Models}",
      journal = {\apj},
         year = 2017,
        month = feb,
       volume = {835},
       number = {2},
          eid = {128},
        pages = {128},
          doi = {10.3847/1538-4357/835/2/128},
archivePrefix = {arXiv},
       eprint = {1604.07824},
 primaryClass = {astro-ph.GA},
       adsurl = {https://ui.adsabs.harvard.edu/abs/2017ApJ...835..128C}
}

@ARTICLE{Romano2021,
       author = {{Romano}, D. and {Magrini}, L. and {Randich}, S. and {Casali}, G. and {Bonifacio}, P. and {Jeffries}, R.~D. and {Matteucci}, F. and {Franciosini}, E. and {Spina}, L. and {Guiglion}, G. and {Chiappini}, C. and {Mucciarelli}, A. and {Ventura}, P. and {Grisoni}, V. and {Bellazzini}, M. and {Bensby}, T. and {Bragaglia}, A. and {de Laverny}, P. and {Korn}, A.~J. and {Martell}, S.~L. and {Tautvai{\v{s}}ien{\.{e}}}, G. and {Carraro}, G. and {Gonneau}, A. and {Jofr{\'e}}, P. and {Pancino}, E. and {Smiljanic}, R. and {Vallenari}, A. and {Fu}, X. and {Guti{\'e}rrez Albarr{\'a}n}, M.~L. and {Jim{\'e}nez-Esteban}, F.~M. and {Montes}, D. and {Damiani}, F. and {Bergemann}, M. and {Worley}, C.},
        title = "{The Gaia-ESO Survey: Galactic evolution of lithium from iDR6}",
      journal = {\aap},
         year = 2021,
        month = sep,
       volume = {653},
          eid = {A72},
        pages = {A72},
          doi = {10.1051/0004-6361/202141340},
archivePrefix = {arXiv},
       eprint = {2106.11614},
 primaryClass = {astro-ph.GA},
       adsurl = {https://ui.adsabs.harvard.edu/abs/2021A&A...653A..72R}
}

@ARTICLE{Beniamini2019,
       author = {{Beniamini}, Paz and {Piran}, Tsvi},
        title = "{The Gravitational waves merger time distribution of binary neutron star systems}",
      journal = {\mnras},
         year = 2019,
        month = aug,
       volume = {487},
       number = {4},
        pages = {4847-4854},
          doi = {10.1093/mnras/stz1589},
archivePrefix = {arXiv},
       eprint = {1903.11614},
 primaryClass = {astro-ph.HE},
       adsurl = {https://ui.adsabs.harvard.edu/abs/2019MNRAS.487.4847B}
}

@ARTICLE{Cote2018b,
       author = {{C{\^o}t{\'e}}, Benoit and {Silvia}, Devin W. and {O'Shea}, Brian W. and {Smith}, Britton and {Wise}, John H.},
        title = "{Validating Semi-analytic Models of High-redshift Galaxy Formation Using Radiation Hydrodynamical Simulations}",
      journal = {\apj},
         year = 2018,
        month = may,
       volume = {859},
       number = {1},
          eid = {67},
        pages = {67},
          doi = {10.3847/1538-4357/aabe8f},
archivePrefix = {arXiv},
       eprint = {1710.06442},
 primaryClass = {astro-ph.GA},
       adsurl = {https://ui.adsabs.harvard.edu/abs/2018ApJ...859...67C}
}

@ARTICLE{Limongi2018,
       author = {{Limongi}, Marco and {Chieffi}, Alessandro},
        title = "{Presupernova Evolution and Explosive Nucleosynthesis of Rotating Massive Stars in the Metallicity Range -3 {\ensuremath{\leq}} [Fe/H] {\ensuremath{\leq}} 0}",
      journal = {\apjs},
         year = 2018,
        month = jul,
       volume = {237},
       number = {1},
          eid = {13},
        pages = {13},
          doi = {10.3847/1538-4365/aacb24},
archivePrefix = {arXiv},
       eprint = {1805.09640},
 primaryClass = {astro-ph.SR},
       adsurl = {https://ui.adsabs.harvard.edu/abs/2018ApJS..237...13L}
}

@ARTICLE{Cristallo2015,
       author = {{Cristallo}, S. and {Straniero}, O. and {Piersanti}, L. and {Gobrecht}, D.},
        title = "{Evolution, Nucleosynthesis, and Yields of AGB Stars at Different Metallicities. III. Intermediate-mass Models, Revised Low-mass Models, and the ph-FRUITY Interface}",
      journal = {\apjs},
         year = 2015,
        month = aug,
       volume = {219},
       number = {2},
          eid = {40},
        pages = {40},
          doi = {10.1088/0067-0049/219/2/40},
archivePrefix = {arXiv},
       eprint = {1507.07338},
 primaryClass = {astro-ph.SR},
       adsurl = {https://ui.adsabs.harvard.edu/abs/2015ApJS..219...40C}
}

@ARTICLE{Nishimura2015,
       author = {{Nishimura}, Nobuya and {Takiwaki}, Tomoya and {Thielemann}, Friedrich-Karl},
        title = "{The r-process Nucleosynthesis in the Various Jet-like Explosions of Magnetorotational Core-collapse Supernovae}",
      journal = {\apj},
         year = 2015,
        month = sep,
       volume = {810},
       number = {2},
          eid = {109},
        pages = {109},
          doi = {10.1088/0004-637X/810/2/109},
archivePrefix = {arXiv},
       eprint = {1501.06567},
 primaryClass = {astro-ph.SR},
       adsurl = {https://ui.adsabs.harvard.edu/abs/2015ApJ...810..109N}
}

@ARTICLE{Paszke2019_pytorch,
       author = {{Paszke}, Adam and {Gross}, Sam and {Massa}, Francisco and {Lerer}, Adam and {Bradbury}, James and {Chanan}, Gregory and {Killeen}, Trevor and {Lin}, Zeming and {Gimelshein}, Natalia and {Antiga}, Luca and {Desmaison}, Alban and {K{\"o}pf}, Andreas and {Yang}, Edward and {DeVito}, Zach and {Raison}, Martin and {Tejani}, Alykhan and {Chilamkurthy}, Sasank and {Steiner}, Benoit and {Fang}, Lu and {Bai}, Junjie and {Chintala}, Soumith},
        title = "{PyTorch: An Imperative Style, High-Performance Deep Learning Library}",
      journal = {arXiv e-prints},
         year = 2019,
        month = dec,
          eid = {arXiv:1912.01703},
        pages = {arXiv:1912.01703},
          doi = {10.48550/arXiv.1912.01703},
archivePrefix = {arXiv},
       eprint = {1912.01703},
 primaryClass = {cs.LG},
       adsurl = {https://ui.adsabs.harvard.edu/abs/2019arXiv191201703P}
}

@ARTICLE{Guiglion2024b,
       author = {{Guiglion}, G. and {Nepal}, S. and {Chiappini}, C. and {Khoperskov}, S. and {Traven}, G. and {Queiroz}, A.~B.~A. and {Steinmetz}, M. and {Valentini}, M. and {Fournier}, Y. and {Vallenari}, A. and {Youakim}, K. and {Bergemann}, M. and {M{\'e}sz{\'a}ros}, S. and {Lucatello}, S. and {Sordo}, R. and {Fabbro}, S. and {Minchev}, I. and {Tautvai{\v{s}}ien{\.{e}}}, G. and {Mikolaitis}, {\v{S}}. and {Montalb{\'a}n}, J.},
        title = "{Beyond Gaia DR3: Tracing the [{\ensuremath{\alpha}}/M] - [M/H] bimodality from the inner to the outer Milky Way disc with Gaia-RVS and convolutional neural networks}",
      journal = {\aap},
         year = 2024,
        month = feb,
       volume = {682},
          eid = {A9},
        pages = {A9},
          doi = {10.1051/0004-6361/202347122},
archivePrefix = {arXiv},
       eprint = {2306.05086},
 primaryClass = {astro-ph.GA},
       adsurl = {https://ui.adsabs.harvard.edu/abs/2024A&A...682A...9G}
}

@ARTICLE{Bergemann2008,
       author = {{Bergemann}, M. and {Gehren}, T.},
        title = "{NLTE abundances of Mn in a sample of metal-poor stars}",
      journal = {\aap},
         year = 2008,
        month = dec,
       volume = {492},
       number = {3},
        pages = {823-831},
          doi = {10.1051/0004-6361:200810098},
archivePrefix = {arXiv},
       eprint = {0811.0681},
 primaryClass = {astro-ph},
       adsurl = {https://ui.adsabs.harvard.edu/abs/2008A&A...492..823B}
}

@ARTICLE{Buder2025,
       author = {{Buder}, Sven and {Kos}, Janez and {Wang}, Xi Ella and {McKenzie}, Madeleine and {Howell}, Madeleine and {Martell}, Sarah and {Hayden}, Michael R. and {Zucker}, Daniel B. and {Nordlander}, Thomas and {Montet}, Benjamin and {Traven}, Gregor and {Bland-Hawthorn}, Joss and {de Silva}, Gayandhi M. and {Freeman}, Kenneth and {Lewis}, Geraint and {Lind}, Karin and {Sharma}, Sanjib and {Simpson}, Jeffrey D. and {Stello}, Dennis and {Zwitter}, Tomaz and {Amarsi}, Anish M. and {Armstrong}, Joseph J. and {Banks}, Kirsten and {Beavis}, Mark and {Beeson}, Kevin-Luke and {Chen}, Boquan and {Ciuc{\u{a}}}, Ioana and {da Costa}, Gary S. and {de Grijs}, Richard and {Martin}, Bailey and {Nataf}, David Moise and {Ness}, Melissa and {Rains}, Adam D. and {Scarr}, Tim and {Vogrin{\v{c}}i{\v{c}}}, Rok and {Wang}, Zixian Purmortal and {Wittenmyer}, Rob A. and {Xie}, Yi Anne and {The Galah Collaboration}},
        title = "{The GALAH survey: Data release 4}",
      journal = {\pasa},
         year = 2025,
        month = may,
       volume = {42},
          eid = {e051},
        pages = {e051},
          doi = {10.1017/pasa.2025.26},
archivePrefix = {arXiv},
       eprint = {2409.19858},
 primaryClass = {astro-ph.GA},
       adsurl = {https://ui.adsabs.harvard.edu/abs/2025PASA...42...51B}
}

@ARTICLE{Rozanski2025,
       author = {{R{\'o}{\.z}a{\'n}ski}, Tomasz and {Ting}, Yuan-Sen and {Jab{\l}o{\'n}ska}, Maja},
        title = "{TransformerPayne: Enhancing Spectral Emulation Accuracy and Data Efficiency by Capturing Long-range Correlations}",
      journal = {\apj},
         year = 2025,
        month = feb,
       volume = {980},
       number = {1},
          eid = {66},
        pages = {66},
          doi = {10.3847/1538-4357/ad9b99},
archivePrefix = {arXiv},
       eprint = {2407.05751},
 primaryClass = {astro-ph.IM},
       adsurl = {https://ui.adsabs.harvard.edu/abs/2025ApJ...980...66R}
}

@ARTICLE{Soubiran2024,
       author = {{Soubiran}, C. and {Creevey}, O.~L. and {Lagarde}, N. and {Brouillet}, N. and {Jofr{\'e}}, P. and {Casamiquela}, L. and {Heiter}, U. and {Aguilera-G{\'o}mez}, C. and {Vitali}, S. and {Worley}, C. and {de Brito Silva}, D.},
        title = "{Gaia FGK benchmark stars: Fundamental T$_{eff}$ and log g of the third version}",
      journal = {\aap},
         year = 2024,
        month = feb,
       volume = {682},
          eid = {A145},
        pages = {A145},
          doi = {10.1051/0004-6361/202347136},
archivePrefix = {arXiv},
       eprint = {2310.11302},
 primaryClass = {astro-ph.SR},
       adsurl = {https://ui.adsabs.harvard.edu/abs/2024A&A...682A.145S}
}

@ARTICLE{Ting2019,
       author = {{Ting}, Yuan-Sen and {Conroy}, Charlie and {Rix}, Hans-Walter and {Cargile}, Phillip},
        title = "{The Payne: Self-consistent ab initio Fitting of Stellar Spectra}",
      journal = {\apj},
         year = 2019,
        month = jul,
       volume = {879},
       number = {2},
          eid = {69},
        pages = {69},
          doi = {10.3847/1538-4357/ab2331},
archivePrefix = {arXiv},
       eprint = {1804.01530},
 primaryClass = {astro-ph.SR},
       adsurl = {https://ui.adsabs.harvard.edu/abs/2019ApJ...879...69T}
}

@ARTICLE{Ryan2000,
       author = {{Ryan}, Sean G. and {Beers}, Timothy C. and {Olive}, Keith A. and {Fields}, Brian D. and {Norris}, John E.},
        title = "{Primordial Lithium and Big Bang Nucleosynthesis}",
      journal = {\apjl},
         year = 2000,
        month = feb,
       volume = {530},
       number = {2},
        pages = {L57-L60},
          doi = {10.1086/312492},
archivePrefix = {arXiv},
       eprint = {astro-ph/9905211},
 primaryClass = {astro-ph},
       adsurl = {https://ui.adsabs.harvard.edu/abs/2000ApJ...530L..57R}
}

@ARTICLE{Pancino2017,
       author = {{Pancino}, E. and {Lardo}, C. and {Altavilla}, G. and {Marinoni}, S. and {Ragaini}, S. and {Cocozza}, G. and {Bellazzini}, M. and {Sabbi}, E. and {Zoccali}, M. and {Donati}, P. and {Heiter}, U. and {Koposov}, S.~E. and {Blomme}, R. and {Morel}, T. and {S{\'\i}mon-D{\'\i}az}, S. and {Lobel}, A. and {Soubiran}, C. and {Montalban}, J. and {Valentini}, M. and {Casey}, A.~R. and {Blanco-Cuaresma}, S. and {Jofr{\'e}}, P. and {Worley}, C.~C. and {Magrini}, L. and {Hourihane}, A. and {Fran{\c{c}}ois}, P. and {Feltzing}, S. and {Gilmore}, G. and {Randich}, S. and {Asplund}, M. and {Bonifacio}, P. and {Drew}, J.~E. and {Jeffries}, R.~D. and {Micela}, G. and {Vallenari}, A. and {Alfaro}, E.~J. and {Allende Prieto}, C. and {Babusiaux}, C. and {Bensby}, T. and {Bragaglia}, A. and {Flaccomio}, E. and {Hambly}, N. and {Korn}, A.~J. and {Lanzafame}, A.~C. and {Smiljanic}, R. and {Van Eck}, S. and {Walton}, N.~A. and {Bayo}, A. and {Carraro}, G. and {Costado}, M.~T. and {Damiani}, F. and {Edvardsson}, B. and {Franciosini}, E. and {Frasca}, A. and {Lewis}, J. and {Monaco}, L. and {Morbidelli}, L. and {Prisinzano}, L. and {Sacco}, G.~G. and {Sbordone}, L. and {Sousa}, S.~G. and {Zaggia}, S. and {Koch}, A.},
        title = "{The Gaia-ESO Survey: Calibration strategy}",
      journal = {\aap},
         year = 2017,
        month = feb,
       volume = {598},
          eid = {A5},
        pages = {A5},
          doi = {10.1051/0004-6361/201629450},
archivePrefix = {arXiv},
       eprint = {1610.06480},
 primaryClass = {astro-ph.SR},
       adsurl = {https://ui.adsabs.harvard.edu/abs/2017A&A...598A...5P}
}

@ARTICLE{Karakas2003,
       author = {{Karakas}, A.~I. and {Lattanzio}, J.~C.},
        title = "{Production of Aluminium and the Heavy Magnesium Isotopes in Asymptotic Giant Branch Stars}",
      journal = {\pasa},
         year = 2003,
        month = jan,
       volume = {20},
       number = {3},
        pages = {279-293},
          doi = {10.1071/AS03010},
       adsurl = {https://ui.adsabs.harvard.edu/abs/2003PASA...20..279K}
}

@ARTICLE{Tolstoy2009,
       author = {{Tolstoy}, Eline and {Hill}, Vanessa and {Tosi}, Monica},
        title = "{Star-Formation Histories, Abundances, and Kinematics of Dwarf Galaxies in the Local Group}",
      journal = {\araa},
         year = 2009,
        month = sep,
       volume = {47},
       number = {1},
        pages = {371-425},
          doi = {10.1146/annurev-astro-082708-101650},
archivePrefix = {arXiv},
       eprint = {0904.4505},
 primaryClass = {astro-ph.CO},
       adsurl = {https://ui.adsabs.harvard.edu/abs/2009ARA&A..47..371T}
}

@ARTICLE{Ness2015,
       author = {{Ness}, M. and {Hogg}, David W. and {Rix}, H. -W. and {Ho}, Anna. Y.~Q. and {Zasowski}, G.},
        title = "{The Cannon: A data-driven approach to Stellar Label Determination}",
      journal = {\apj},
         year = 2015,
        month = jul,
       volume = {808},
       number = {1},
          eid = {16},
        pages = {16},
          doi = {10.1088/0004-637X/808/1/16},
archivePrefix = {arXiv},
       eprint = {1501.07604},
 primaryClass = {astro-ph.SR},
       adsurl = {https://ui.adsabs.harvard.edu/abs/2015ApJ...808...16N}
}

@ARTICLE{Gehren2004,
       author = {{Gehren}, T. and {Liang}, Y.~C. and {Shi}, J.~R. and {Zhang}, H.~W. and {Zhao}, G.},
        title = "{Abundances of Na, Mg and Al in nearby metal-poor stars}",
      journal = {\aap},
         year = 2004,
        month = jan,
       volume = {413},
        pages = {1045-1063},
          doi = {10.1051/0004-6361:20031582},
       adsurl = {https://ui.adsabs.harvard.edu/abs/2004A&A...413.1045G}
}

@ARTICLE{Gehren2006,
       author = {{Gehren}, T. and {Shi}, J.~R. and {Zhang}, H.~W. and {Zhao}, G. and {Korn}, A.~J.},
        title = "{Na, Mg and Al abundances as a population discriminant for nearby metal-poor stars}",
      journal = {\aap},
         year = 2006,
        month = jun,
       volume = {451},
       number = {3},
        pages = {1065-1079},
          doi = {10.1051/0004-6361:20054434},
       adsurl = {https://ui.adsabs.harvard.edu/abs/2006A&A...451.1065G}
}

@ARTICLE{Heiter2015b,
       author = {{Heiter}, U. and {Jofr{\'e}}, P. and {Gustafsson}, B. and {Korn}, A.~J. and {Soubiran}, C. and {Th{\'e}venin}, F.},
        title = "{Gaia FGK benchmark stars: Effective temperatures and surface gravities}",
      journal = {\aap},
         year = 2015,
        month = oct,
       volume = {582},
          eid = {A49},
        pages = {A49},
          doi = {10.1051/0004-6361/201526319},
archivePrefix = {arXiv},
       eprint = {1506.06095},
 primaryClass = {astro-ph.SR},
       adsurl = {https://ui.adsabs.harvard.edu/abs/2015A&A...582A..49H}
}

@ARTICLE{Venn2004,
       author = {{Venn}, Kim A. and {Irwin}, Mike and {Shetrone}, Matthew D. and {Tout}, Christopher A. and {Hill}, Vanessa and {Tolstoy}, Eline},
        title = "{Stellar Chemical Signatures and Hierarchical Galaxy Formation}",
      journal = {\aj},
         year = 2004,
        month = sep,
       volume = {128},
       number = {3},
        pages = {1177-1195},
          doi = {10.1086/422734},
archivePrefix = {arXiv},
       eprint = {astro-ph/0406120},
 primaryClass = {astro-ph},
       adsurl = {https://ui.adsabs.harvard.edu/abs/2004AJ....128.1177V}
}

@ARTICLE{Storm2025,
       author = {{Storm}, Nicholas and {Bergemann}, Maria and {Eitner}, Philipp and {Hoppe}, Richard and {Kemp}, Alex J. and {Ruiter}, Ashley J. and {Janka}, Hans-Thomas and {Sieverding}, Andre and {de Mink}, Selma E. and {Seitenzahl}, Ivo R. and {Owusu}, Evans K.},
        title = "{Observational constraints on the origin of the elements. IX. 3D NLTE abundances of metals in the context of Galactic Chemical Evolution models and 4MOST}",
      journal = {\mnras},
         year = 2025,
        month = apr,
       volume = {538},
       number = {4},
        pages = {3284-3313},
          doi = {10.1093/mnras/staf472},
archivePrefix = {arXiv},
       eprint = {2503.16946},
 primaryClass = {astro-ph.SR},
       adsurl = {https://ui.adsabs.harvard.edu/abs/2025MNRAS.538.3284S}
}

@ARTICLE{Hegedus2025,
       author = {{Heged{\H{u}}s}, Viola and {M{\'e}sz{\'a}ros}, Szabolcs and {Vil{\'a}gos}, Blanka and {Pignatari}, Marco and {Griffith}, Emily J. and {Souto}, Diogo and {Lugaro}, Maria},
        title = "{Reconstructing the Milky Way chemical map with the galactic chemical evolution tool OMEGA+ from SDSS-MWM}",
      journal = {\aap},
         year = 2025,
        month = jul,
       volume = {699},
          eid = {A293},
        pages = {A293},
          doi = {10.1051/0004-6361/202553951},
archivePrefix = {arXiv},
       eprint = {2506.00503},
 primaryClass = {astro-ph.GA},
       adsurl = {https://ui.adsabs.harvard.edu/abs/2025A&A...699A.293H}
}

@ARTICLE{deJong2019,
       author = {{de Jong}, R.~S. and {Agertz}, O. and {Berbel}, A.~A. and {Aird}, J. and {Alexander}, D.~A. and {Amarsi}, A. and {Anders}, F. and {Andrae}, R. and {Ansarinejad}, B. and {Ansorge}, W. and {Antilogus}, P. and {Anwand-Heerwart}, H. and {Arentsen}, A. and {Arnadottir}, A. and {Asplund}, M. and {Auger}, M. and {Azais}, N. and {Baade}, D. and {Baker}, G. and {Baker}, S. and {Balbinot}, E. and {Baldry}, I.~K. and {Banerji}, M. and {Barden}, S. and {Barklem}, P. and {Barth{\'e}l{\'e}my-Mazot}, E. and {Battistini}, C. and {Bauer}, S. and {Bell}, C.~P.~M. and {Bellido-Tirado}, O. and {Bellstedt}, S. and {Belokurov}, V. and {Bensby}, T. and {Bergemann}, M. and {Bestenlehner}, J.~M. and {Bielby}, R. and {Bilicki}, M. and {Blake}, C. and {Bland-Hawthorn}, J. and {Boeche}, C. and {Boland}, W. and {Boller}, T. and {Bongard}, S. and {Bongiorno}, A. and {Bonifacio}, P. and {Boudon}, D. and {Brooks}, D. and {Brown}, M.~J.~I. and {Brown}, R. and {Br{\"u}ggen}, M. and {Brynnel}, J. and {Brzeski}, J. and {Buchert}, T. and {Buschkamp}, P. and {Caffau}, E. and {Caillier}, P. and {Carrick}, J. and {Casagrande}, L. and {Case}, S. and {Casey}, A. and {Cesarini}, I. and {Cescutti}, G. and {Chapuis}, D. and {Chiappini}, C. and {Childress}, M. and {Christlieb}, N. and {Church}, R. and {Cioni}, M. -R.~L. and {Cluver}, M. and {Colless}, M. and {Collett}, T. and {Comparat}, J. and {Cooper}, A. and {Couch}, W. and {Courbin}, F. and {Croom}, S. and {Croton}, D. and {Daguis{\'e}}, E. and {Dalton}, G. and {Davies}, L.~J.~M. and {Davis}, T. and {de Laverny}, P. and {Deason}, A. and {Dionies}, F. and {Disseau}, K. and {Doel}, P. and {D{\"o}scher}, D. and {Driver}, S.~P. and {Dwelly}, T. and {Eckert}, D. and {Edge}, A. and {Edvardsson}, B. and {Youssoufi}, D.~E. and {Elhaddad}, A. and {Enke}, H. and {Erfanianfar}, G. and {Farrell}, T. and {Fechner}, T. and {Feiz}, C. and {Feltzing}, S. and {Ferreras}, I. and {Feuerstein}, D. and {Feuillet}, D. and {Finoguenov}, A. and {Ford}, D. and {Fotopoulou}, S. and {Fouesneau}, M. and {Frenk}, C. and {Frey}, S. and {Gaessler}, W. and {Geier}, S. and {Gentile Fusillo}, N. and {Gerhard}, O. and {Giannantonio}, T. and {Giannone}, D. and {Gibson}, B. and {Gillingham}, P. and {Gonz{\'a}lez-Fern{\'a}ndez}, C. and {Gonzalez-Solares}, E. and {Gottloeber}, S. and {Gould}, A. and {Grebel}, E.~K. and {Gueguen}, A. and {Guiglion}, G. and {Haehnelt}, M. and {Hahn}, T. and {Hansen}, C.~J. and {Hartman}, H. and {Hauptner}, K. and {Hawkins}, K. and {Haynes}, D. and {Haynes}, R. and {Heiter}, U. and {Helmi}, A. and {Aguayo}, C.~H. and {Hewett}, P. and {Hinton}, S. and {Hobbs}, D. and {Hoenig}, S. and {Hofman}, D. and {Hook}, I. and {Hopgood}, J. and {Hopkins}, A. and {Hourihane}, A. and {Howes}, L. and {Howlett}, C. and {Huet}, T. and {Irwin}, M. and {Iwert}, O. and {Jablonka}, P. and {Jahn}, T. and {Jahnke}, K. and {Jarno}, A. and {Jin}, S. and {Jofre}, P. and {Johl}, D. and {Jones}, D. and {J{\"o}nsson}, H. and {Jordan}, C. and {Karovicova}, I. and {Khalatyan}, A. and {Kelz}, A. and {Kennicutt}, R. and {King}, D. and {Kitaura}, F. and {Klar}, J. and {Klauser}, U. and {Kneib}, J. -P. and {Koch}, A. and {Koposov}, S. and {Kordopatis}, G. and {Korn}, A. and {Kosmalski}, J. and {Kotak}, R. and {Kovalev}, M. and {Kreckel}, K. and {Kripak}, Y. and {Krumpe}, M. and {Kuijken}, K. and {Kunder}, A. and {Kushniruk}, I. and {Lam}, M.~I. and {Lamer}, G. and {Laurent}, F. and {Lawrence}, J. and {Lehmitz}, M. and {Lemasle}, B. and {Lewis}, J. and {Li}, B. and {Lidman}, C. and {Lind}, K. and {Liske}, J. and {Lizon}, J. -L. and {Loveday}, J. and {Ludwig}, H. -G. and {McDermid}, R.~M. and {Maguire}, K. and {Mainieri}, V. and {Mali}, S. and {Mandel}, H.},
        title = "{4MOST: Project overview and information for the First Call for Proposals}",
      journal = {The Messenger},
         year = 2019,
        month = mar,
       volume = {175},
        pages = {3-11},
          doi = {10.18727/0722-6691/5117},
archivePrefix = {arXiv},
       eprint = {1903.02464},
 primaryClass = {astro-ph.IM},
       adsurl = {https://ui.adsabs.harvard.edu/abs/2019Msngr.175....3D}
}

@ARTICLE{Fuhrmann1998,
       author = {{Fuhrmann}, Klaus},
        title = "{Nearby stars of the Galactic disk and halo}",
      journal = {\aap},
         year = 1998,
        month = oct,
       volume = {338},
        pages = {161-183},
       adsurl = {https://ui.adsabs.harvard.edu/abs/1998A&A...338..161F}
}

@PHDTHESIS{Sneden1973,
       author = {{Sneden}, Christopher Alan},
        title = "{Carbon and Nitrogen Abundances in Metal-Poor Stars.}",
       school = {University of Texas, Austin},
         year = 1973,
        month = jan,
       adsurl = {https://ui.adsabs.harvard.edu/abs/1973PhDT.......180S}
}

@misc{Reetz1991,
    author = {{Reetz}, J.~K.},
    title = "{Diploma Thesis, Universität München}",    year = 1991
}

@ARTICLE{Wheeler2023,
       author = {{Wheeler}, Adam J. and {Abruzzo}, Matthew W. and {Casey}, Andrew R. and {Ness}, Melissa K.},
        title = "{KORG: A Modern 1D LTE Spectral Synthesis Package}",
      journal = {\aj},
         year = 2023,
        month = feb,
       volume = {165},
       number = {2},
          eid = {68},
        pages = {68},
          doi = {10.3847/1538-3881/acaaad},
archivePrefix = {arXiv},
       eprint = {2211.00029},
 primaryClass = {astro-ph.SR},
       adsurl = {https://ui.adsabs.harvard.edu/abs/2023AJ....165...68W}
}

@ARTICLE{Wehrhahn2023,
       author = {{Wehrhahn}, A. and {Piskunov}, N. and {Ryabchikova}, T.},
        title = "{PySME. Spectroscopy Made Easier}",
      journal = {\aap},
         year = 2023,
        month = mar,
       volume = {671},
          eid = {A171},
        pages = {A171},
          doi = {10.1051/0004-6361/202244482},
archivePrefix = {arXiv},
       eprint = {2210.04755},
 primaryClass = {astro-ph.SR},
       adsurl = {https://ui.adsabs.harvard.edu/abs/2023A&A...671A.171W}
}

@ARTICLE{Wedemeyer2017,
       author = {{Wedemeyer}, Sven and {Ku{\v{c}}inskas}, Ar{\={u}}nas and {Klevas}, Jonas and {Ludwig}, Hans-G{\"u}nter},
        title = "{Three-dimensional hydrodynamical CO$^{5}$BOLD model atmospheres of red giant stars. VI. First chromosphere model of a late-type giant}",
      journal = {\aap},
         year = 2017,
        month = oct,
       volume = {606},
          eid = {A26},
        pages = {A26},
          doi = {10.1051/0004-6361/201730405},
archivePrefix = {arXiv},
       eprint = {1705.09641},
 primaryClass = {astro-ph.SR},
       adsurl = {https://ui.adsabs.harvard.edu/abs/2017A&A...606A..26W}
}
\bibliographystyle{aasjournalv7}

\appendix
\restartappendixnumbering

\section{Details on the neural network training}
\label{app:payne_training}

\begin{figure}[ht]
\includegraphics[width=1.0\columnwidth]{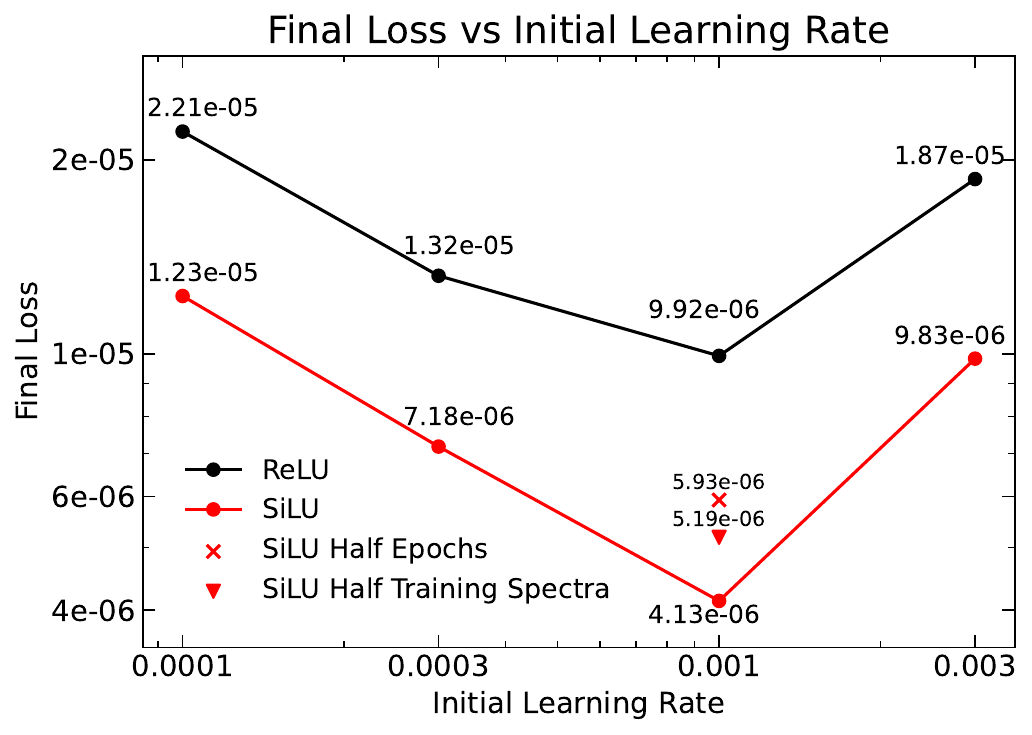}
\caption{Final validation loss of the neural networks as a function of the initial learning rate. The neural networks indicated as top black points were trained using ReLU activation functions, while the bottom red ones used SiLU instead. Except for the two separate activation functions and the initial learning rate, all of the networks were trained in an identical manner.
\label{fig:loss}}
\end{figure}

In this section, we outline the technical details of our ANN training method. The Payne was implemented and trained using PyTorch \citep{Paszke2019_pytorch} and consists of 3 fully-connected hidden layers of 1024 neurons each, using Sigmoid Linear Unit (SiLU) activation functions, and a final output layer of size 33375 with a sigmoid activation function. The latter limits the output to within 0 to 1, consistent with the range of normalised spectra. The network size was chosen as a balance between complexity and the speed of the network \citep[e.g. see Fig. 4 in][]{Rozanski2025}. Figure~\ref{fig:loss} shows a summary of our optimization experiments that consisted of training a series of networks on a smaller training set of $288\,211$ spectra. In particular, we trained networks with two different configurations -- one using Rectified Linear Unit (ReLU) activation functions (black) and one that employed SiLU instead (red) --  for a variety of initial learning rates between 0.0001 and 0.003. \ns{ReLU refers to a function ${\rm ReLU}(x) = \max(0, x)$, while SiLU is a smoother approximation ${\rm SiLU}(x) = x \times {\rm sigmoid}(x)$.} For SiLU we also tested a network trained for half the steps (red cross) and using only half of the training spectra (red triangle). The main take-away is that a right choice of an activation function reduces the final loss significantly, in this case by around half. \ns{This is not unexpected, since SiLU is smoother than ReLU. This is the typical behaviour we expect to see in stellar spectra - e.g. absorption lines become weaker and do not disappear in a step-function manner.} We also did a brief test, skipping the final sigmoid layer \ns{after the output}. This significantly increased the final loss by around a factor of 2. A too low or too high initial learning rate can also result in a suboptimal training convergence. For our network, reducing the number of training steps or training spectra by half had the smallest impact. Out of the final training set of $404~793$ spectra, 6\% were used for the validation set. The individual abundance values were chosen in a uniformly random fashion, thus the network a priori contains no elemental abundance bias and thus should be able to fit non-standard chemical composition stars. Our training spectra did not contain any broadening. However, it was dynamically applied after the emulation from Payne during fitting, similar to how it is done with classical stellar synthesis codes. This simplifies the network, reduces the amount of training spectra needed, and thus improves the convergence. The network was trained on 4 A100 GPUs. The loss was calculated using mean squared difference, AdamW optimiser with a cosine-type scheduler, which dynamically reduced learning rate over time for better network convergence. 

We did not highlight this in the figure, but the network size has a direct impact as well. We tried both smaller (256 neurons per hidden layer) and larger (2048 neurons per hidden layer) networks. In our case, a larger network did not significantly decrease the final loss - by only around 20\%. However, a smaller network had a bigger impact - 4.2 times smaller network had 3.8 times larger loss (256 neurons per hidden layer resulted in a loss of $1.56 \times 10^{-5}$). We also tested the performance of the smaller network on the stellar parameters and abundances. Stellar parameters and most of abundances had a nearly identical recovery rate as the big network, with only slight increase in bias and spread compared to TSFitPy values (typically by around 0.02 dex). However, both Y and Eu bias suffered significantly, especially at the low abundance ranges. For europium, the average bias went to 0.3 dex (up to 0.7-1.0 dex for the lowest abundances). While the effect of the larger loss was not seen for most elements, it is more noticeable for weaker lines. Thus we decided to stick with the network that has 1024 neurons per hidden layer, as the balance between inference time and accuracy of the network. 

The final technical detail that is relevant for this network architecture is the potential to decrease inference time. Let's imagine that one needs to fit only one specific line in the spectrum. In this case, skipping synthesis of the remaining parts of the spectrum could speed up the process. This is possible for our Payne network. Since the network is constructed as an input layer connected with hidden layers and the final output layer, one requires only to do all of the initial matrix multiplications first. The calculation in the output layer can skip all of the pixels except for the ones that are of interest. The last matrix is the largest one by far ($1024 \times 33375$ in our case, compared to the $1024 \times 1024$ and $21 \times 1024$ for other ones), so skipping calculations in the final matrix can give a significant time gain, despite using a ``big'' Payne network.

\section{Validation of the network performance}

\begin{figure*}[ht!]
\includegraphics[width=1.0\textwidth]{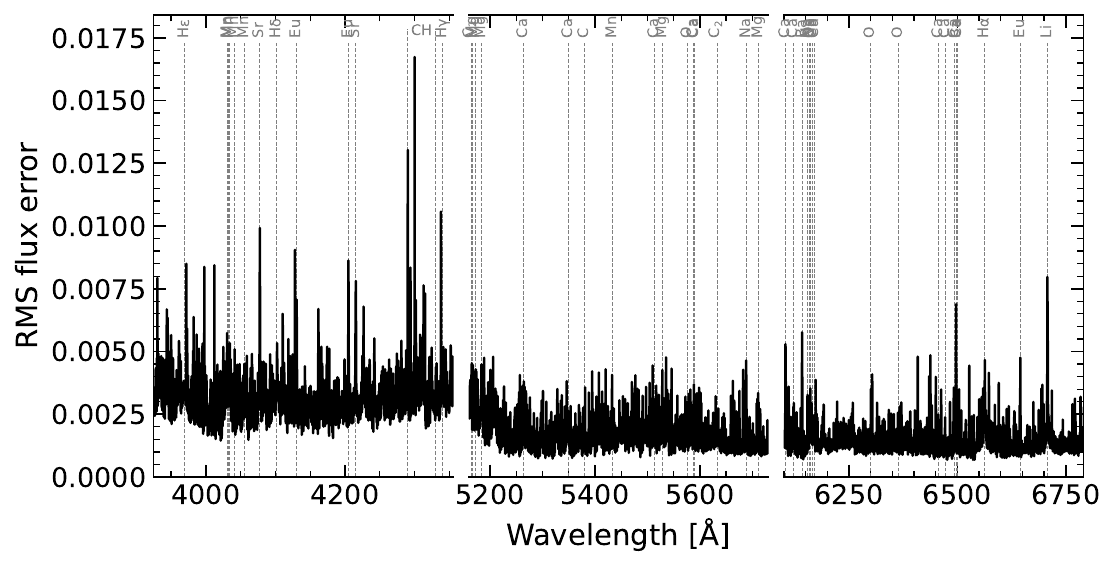}
\caption{\ns{Mean RMS normalised flux error across the stellar spectrum range for all validation synthetic spectra. Some of the notables lines are overplotted as well.}
\label{fig:rmse_per_pixel}}
\end{figure*}

\begin{figure}
\includegraphics[width=1\columnwidth]{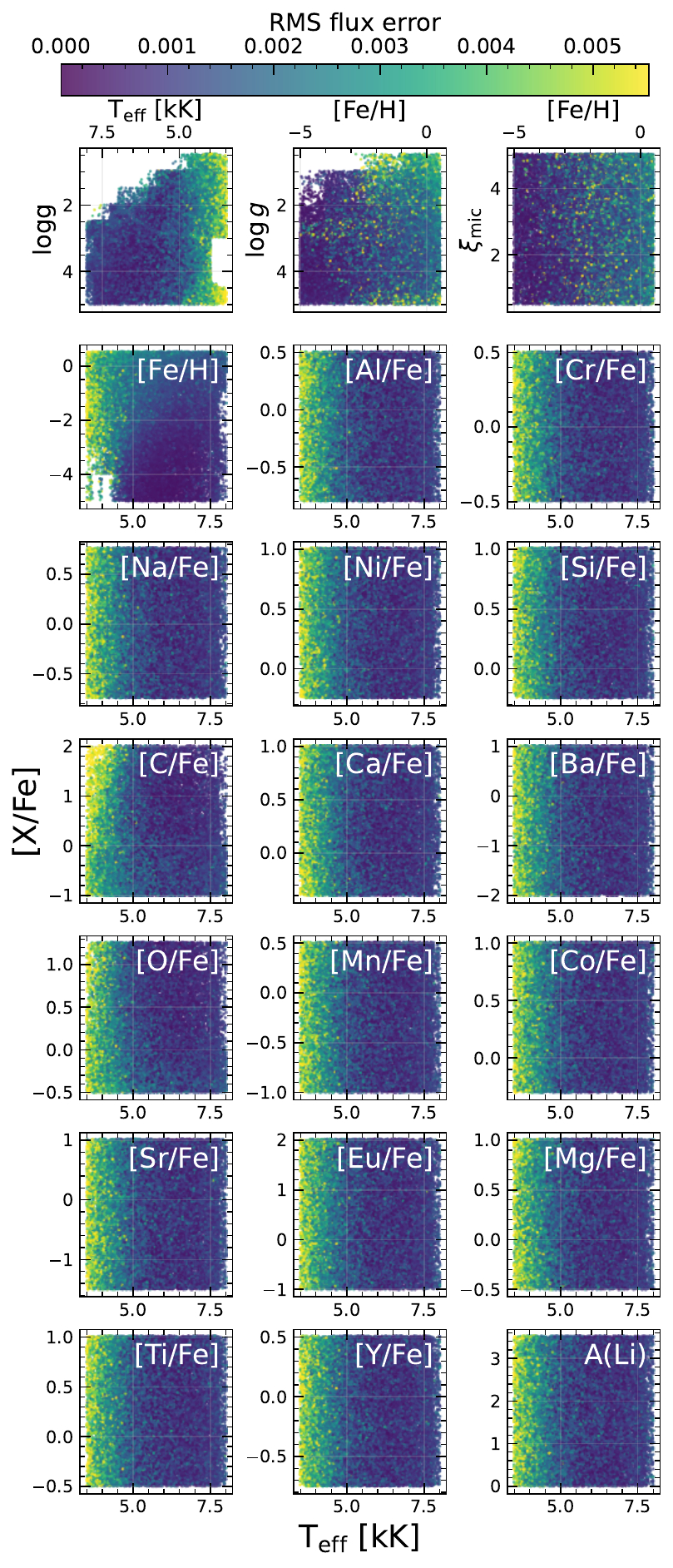}
\caption{\ns{Distribution of mean root mean square (RMS) flux error of each spectrum (averaged across wavelength) in a 2D histogram of validation set of synthetic spectra. Top left panel shows the $\teff$-$\logg$ diagram, the other two top panels are distributions as a function of [Fe/H], and the rest are as a function of $\teff$. This distribution is shown because $\teff$ has the largest impact on the RMS error compared to any parameter in the grid.}
\label{fig:rmse_parameter_space}}
\end{figure}

\begin{figure}
\includegraphics[width=1\columnwidth]{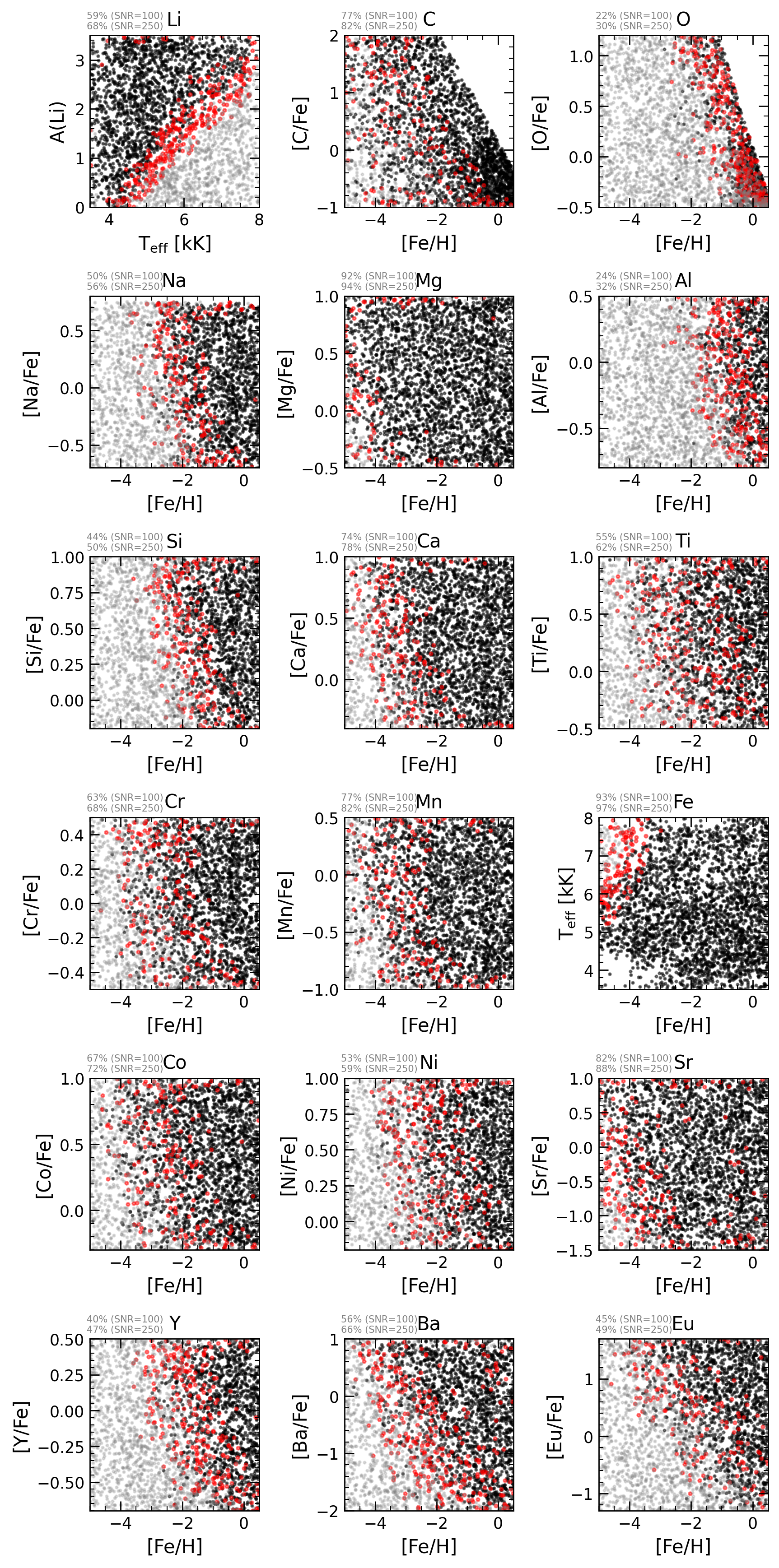}
\caption{\ns{Recovery rate of the true abundances for the fraction of the validation set of synthetic spectra with added noise (SNR = 100 and 250 per \AA). Black points indicate abundances recovered at least for SNR = 100 per \AA, red points indicate abundances recovered for SNR = 250 per \AA, and grey points indicate non-recovered abundances. The number in the top left of each subplot indicates the total recovery percentage.}
\label{fig:recovery_rate_hr}}
\end{figure}

\ns{We also tested the performance of the network across the parameter space. Fig. \ref{fig:rmse_per_pixel} shows mean RMS normalised flux error for the validation synthetic spectra set across the full parameter space. The RMS error was calculated by using input parameters for each synthetic spectrum from the validation set, producing a Payne-equivalent spectrum and calculating an error compared to the original input Turbospectrum model using:}

\begin{equation}
    \mathrm{RMS~error} = \sqrt{\frac{1}{P}\sum_{j=1}^{P} \left(\hat y_{j}-y_{j}\right)^2 },
\end{equation}

\ns{where $P$ is the number of pixels, $\hat y_{j}$ is the predicted spectrum from Payne, $y_{j}$ is the input spectrum from Turbospectrum and $j$ is each pixel. On average, the blue window shows the largest error, due to dominance of molecular bands. A large error is present in the cores of the CH G-band at $\approx 4300$ \AA~lines, but also in the cores of H~I and Li~I lines. On average, the error is $<0.005$ (less than 0.5 percent) in normalised flux units, which implies that the error associated with the Payne interpolation is also typically $<0.005$ flux units. }

\ns{Fig. \ref{fig:rmse_parameter_space} shows the mean RMS flux error across each spectrum for the full parameter space range. Based on the $\teff$-$\logg$ diagram, lower $\teff$ values result in higher RMS error, because of the stronger molecular lines. Thus the spectra with $\teff > 5000$ K have mean RMS error $< 0.002$, also implying that lower [Fe/H] spectra at the same $\teff$ have a smaller RMS error. Typically, individual abundance values do not have an effect on the RMS error, except for the abundance of carbon. [C/Fe] $\approx 2$ at the lowest $\teff$ results in the highest RMS error values.}

\ns{In Fig. \ref{fig:recovery_rate_hr} we show the recovery rate of true abundances for a fraction of synthetic spectra in the validation set with added noise (SNR = 100 and 250 per \AA). Black points indicate that the abundance was recovered in spectra with at least SNR = 100 per \AA, red points were recovered only for SNR = 250 per \AA, while grey ones indicate spectra with non-recovered abundance. This shows the best possible performance limits of the current fitting with this network, although we note that a different linemask selection or fitting algorithm would change the distribution. Overall the results are encouraging: metallicity (parametrized by [Fe/H]) is recovered in almost all spectra, except for the lowest metallicity regime in the hottest stars at lower SNR. Each element group shows a good recovery rate across all metallicity regime: Mg for $\alpha$-elements, Mn for iron-peak elements and Sr for neutron-capture elements. Li recovery depends directly on the stellar effective temperature, while O and Al have the worst recovery properties due to the weakness of their spectral lines.}

\section{\ns{Results for the solar spectra}}
\label{app:solar_spectra}

\ns{In Tab. \ref{tab:sun_corr} we present the average abundances obtained using five high-resolution solar spectra (NARVAL, UVES and HARPS spectrographs), which we use to self-consistently compute the stellar abundances in the square bracket notation. The solar abundance of Y is not very accurate, possibly either because of the use of older sources of $\log gf$ values \citep[see also discussion in Sec. 2.2. in][]{Storm2024} or due to the lack of strong reliable diagnostic lines in 4MOST-HR spectral range.}

\begin{deluxetable}{cc}[!htbp]
\tabletypesize{\scriptsize}
\digitalasset
\tablewidth{0pt}
\tablecaption{\ns{Average fitted solar abundances based on five 4MOSTified solar spectra.}}
\label{tab:sun_corr}
\tablehead{
\colhead{Element} & \colhead{Fitted abundance}
}
\startdata
A(Al) & 6.45 \\
A(Ba) & 2.15 \\
A(C) &  8.43 \\
A(Ca) & 6.43 \\
A(Co) & 4.96 \\
A(Cr) & 5.65 \\
A(Eu) & 0.42 \\
A(Fe) & 7.45 \\
A(Mg) & 7.53 \\
A(Mn) & 5.44 \\
A(Na) & 6.29 \\
A(Ni) & 6.31 \\
A(O) &  8.74 \\
A(Si) & 7.50 \\
A(Ti) & 5.08 \\
A(Y) &  1.98 \\
\enddata
\end{deluxetable}

\section{Lines used for fitting}
\label{app:linemask}

\begin{deluxetable*}{cccccc}
\digitalasset
\tablecaption{Table with the atomic data of the lines used in the fitting of different elements \label{tab:lines}. \ns{The full table will be made available in machine-readable format via CDS upon publication of the paper.}}
\tablehead{
\colhead{Element} & \colhead{$\lambda$ [\AA]} & \colhead{$E_{low}$ [eV]} & \colhead{$\log(gf)$}& \colhead{$g_{up}$}& \colhead{Isotope (if any)}}
\startdata
Al I & 6696.023 & 3.143 & -1.569 & 4 &  \\
Al I & 6698.673 & 3.143 & -1.870 & 2 &  \\
Ba II & 6141.709 & 0.704 & -0.395 & 4 & 137 \\
Ba II & 6141.713 & 0.704 & -0.032 & 4 & 134 \\
Ba II & 6141.714 & 0.704 & -0.032 & 4 & 136 \\
\enddata
\end{deluxetable*}

\section{Fitted parameters for all spectra}
\begin{deluxetable*}{ccccccccc}
\tabletypesize{\scriptsize}
\digitalasset
\tablewidth{0pt}
\tablecaption{Fitted parameters for all spectra. The uncertainty includes both systematic and errors returned by \ns{\texttt{LMFIT}}. Instruments: $^1$ - UVES, $^2$ - NARVAL, $^3$ - UVES-POP, $^4$ - HARPS, $^5$ - FOCES. \ns{The full table will be made available in machine-readable format via CDS upon publication of the paper.}}
\label{tab:fitted_param1}
\tablehead{
Star&$\teff$& $\logg$ & $\vmic$ & [Fe/H]& $v_{\text{brd}}$  &[Al/Fe] & [Cr/Fe] & [Na/Fe]   \\
& [K]&    & [km s$^{-1}$]&  & [km s$^{-1}$]&  &
}
\startdata
15PEG$^5$ & $6488 \pm 137$ & $4.11 \pm 0.25$ & $1.58 \pm 0.01$ & $-0.47 \pm 0.06$  & $11.18 \pm 0.03$ & -& $-0.16 \pm 0.13$& $0.10 \pm 0.08$ \\
18Sco$^1$ & $5828 \pm 135$ & $4.49 \pm 0.25$ & $1.36 \pm 0.01$ & $-0.05 \pm 0.06$  & $1.39 \pm 0.02$ & $0.12 \pm 0.07$& $-0.06 \pm 0.11$& $0.02 \pm 0.08$ \\
18Sco$^2$ & $5835 \pm 134$ & $4.52 \pm 0.24$ & $1.23 \pm 0.00$ & $-0.04 \pm 0.06$  & $1.41 \pm 0.01$ & $0.10 \pm 0.07$& $-0.05 \pm 0.10$& $0.01 \pm 0.07$ \\
31AQL$^5$ & $5575 \pm 136$ & $4.23 \pm 0.25$ & $1.10 \pm 0.01$ & $0.34 \pm 0.06$  & $4.26 \pm 0.04$ & $0.12 \pm 0.08$& $-0.01 \pm 0.18$& $0.19 \pm 0.08$ \\
Arcturus$^2$ & $4165 \pm 134$ & $1.30 \pm 0.25$ & $1.58 \pm 0.00$ & $-0.72 \pm 0.06$  & $2.84 \pm 0.02$ & $0.35 \pm 0.07$& $0.05 \pm 0.11$& $0.14 \pm 0.07$ \\
\enddata
\end{deluxetable*}

\end{document}